\documentclass[12pt,aps,prd,onecolumn,floatfix, tightenlines,showpacs,showkeys,preprintnumbers, longbibliography]{revtex4-1}
\pdfoutput=1
\usepackage[colorlinks=true,citecolor=blue,linkcolor=blue,breaklinks=true]{hyperref}
\usepackage{epsfig}
\usepackage{hyperref} 
\usepackage{graphicx}
\usepackage{slashed}    
\usepackage{bm}  
\usepackage{multirow}  
\usepackage{color}  
\usepackage{soul}
\newcommand{\be}{\begin{eqnarray}}  
\newcommand{\ee}{\end{eqnarray}}

\begin{document}  

\title{Model-independent diagnostic of self-induced spectral equalization versus ordinary matter effects in supernova neutrinos}

\author{Francesco Capozzi}
\email{capozzi@mppmu.mpg.de}
\affiliation{Max Planck Institut f\"{u}r Physik (Werner-Heisenberg-Institut), F\"{o}hringer Ring 6, 80805 M\"{u}nchen, Germany. 
}   
          
\author{Basudeb Dasgupta}
\email{bdasgupta@theory.tifr.res.in}
\affiliation{Tata Institute of Fundamental Research,
             Homi Bhabha Road, Mumbai 400005, India.}
\author{Alessandro Mirizzi}
\email{alessandro.mirizzi@ba.infn.it }
\affiliation{Dipartimento Interateneo di Fisica ``Michelangelo Merlin'', Via Amendola 173, 70126 Bari, Italy.}
\affiliation{Istituto Nazionale di Fisica Nucleare - Sezione di Bari,
Via Amendola 173, 70126 Bari, Italy.}

\begin{abstract}   
Self-induced flavor conversions near the supernova (SN) core can make the fluxes for different neutrino species become almost equal, potentially altering the dynamics of the SN explosion and washing out all further neutrino oscillation effects. We present a new model-independent analysis strategy for the next galactic SN signal that will distinguish this flavor equalization scenario from a matter effects only scenario during the SN accretion phase. Our method does not rely on fitting or modelling the energy-dependent fluences of the different species to a known function, but rather uses a model-independent comparison of charged-current and neutral-current  events at large next-generation underground detectors. Specifically,  we advocate that the events due to elastic scattering on protons in a scintillator detector, which is insensitive to oscillation effects and can be used as a model-independent normalization, should be compared with the events due to inverse beta decay of $\bar\nu_e$ in a water Cherenkov detector and/or the events due to charged-current interactions of $\nu_e$ in an Argon detector. The ratio of events in these different detection channels allow one to distinguish a complete flavor equalization from a pure matter effect, for either of the neutrino mass orderings, as long as the spectral differences among the different species are not too small.
\end{abstract}   
\preprint{TIFR/TH/18-15}  
\preprint{MPP-2018-147}

\maketitle

\section{Introduction}         \label{intro}  

The detection of the next galactic supernova (SN) neutrino burst
stands out as one of the next frontiers of low-energy neutrino astronomy.
It is expected that such an event will lead to an immense improvement
in our understanding of the SN explosion dynamics and neutrino flavor 
mixing (see~\cite{Duan:2010bg,Horiuchi:2017sku,Mirizzi:2015eza,Scholberg:2012id}  for recent reviews). 
In this context,  remarkable attention has been devoted to the observable signatures associated with the Mikheyev-Smirnov-Wolfenstein (MSW) matter effect~\cite{Wolfenstein:1977ue,Mikheev:1986gs} on the neutrino flavor evolution in SNe~\mbox{\cite{Dighe:1999bi,Fogli:2004ff, Lunardini:2003eh,Dasgupta:2005wn}}.
On the other hand, it has now been known for over a decade that the matter effects are not the only source of
flavor conversions in SNe. In the deepest regions inside a SN, the flavor evolution is determined
by neutrino-neutrino forward scattering that can induce large flavor conversions~\cite{Duan:2006an,Hannestad:2006nj,Fogli:2007bk,Dasgupta:2009mg}. Despite the frenetic attempts 
 to characterize these effects (see~\cite{Duan:2010bg, Mirizzi:2015eza,Chakraborty:2016lct} for reviews), their current description is still far from being settled.
In some situations they are expected to induce spectral swaps and splits that would be further modified by the matter effects~\cite{Duan:2006an,Fogli:2007bk,Dasgupta:2009mg,Choubey:2010up}. While in other cases,
they may lead to flavor decoherence leading to equalization of fluxes and spectra for the different neutrino species~\cite{Sawyer:2008zs,Sawyer:2015dsa}.

It is worthwhile to ask if there is a time window in which oscillations effects are more prominent and the different flavor conversion scenarios can be observationally distinguished. Oscillation effects on the $\nu_e$ neutronization burst, occurring a few milliseconds after the core-bounce, have a rather clear interpretation. Due to the large excess of $\nu_e$ over the other species, the self-induced oscillations would be suppressed and the $\nu_e$ flux would be processed by only the matter effects~\cite{Hannestad:2006nj}. Even if one cannot extract information on self-induced effects during the neutronization burst,  the (non)observation of the associated $\nu_e$ peak would probe the unknown neutrino mass ordering~\cite{Kachelriess:2004ds}.  In the subsequent accretion phase,  lasting until post-bounce times $t_{\rm pb}\lesssim 0.5$~s,  (anti)neutrinos of all the flavors are copiously emitted, with a  sizeable flavor hierarchy between electron and non-electron neutrino spectra. Given this large flux hierarchy, oscillation effects would be prominent during this phase.
Finally, during the cooling phase, i.e., for post-bounce times $t_{\rm pb}\gtrsim 1$~s, recent supernova simulations indicate that neutrino fluxes have only a modest flavor-dependence. Thus, later times are likely to be less promising for observing flavor conversion signals~\cite{Mirizzi:2015eza}. In a nutshell, one finds that  both neutronization burst and accretion phase are promising for studying flavor conversion, but signatures of self-induced conversions are more likely in the accretion phase.  

During the accretion phase, neutrinos may either show only the matter effect driven flavor conversions, or the self-induced conversions as well, that may lead to flavor equalization. It is not known which of these scenarios is borne out in Nature. For iron core SNe, the matter potential is expected to dominate the neutrino-neutrino potential, and thus strongly suppress the self-induced effects~\cite{Chakraborty:2011nf,Sarikas:2011am}. If this is the case, the neutrino flavor evolution is determined by matter effects only. However, the above-said suppression has been recently questioned. It has been shown that the presence of temporal instabilities of the dense neutrino gas allow for self-induced effects, even in the presence of a dominant matter density~\cite{Dasgupta:2015iia}.  Also,  fast flavor conversions just above  the SN core may not be inhibited  by a large matter density~\cite{Sawyer:2015dsa,Chakraborty:2016lct,Dasgupta:2016dbv}. All these effects go in the same direction: an equalization of neutrino fluxes of different neutrinos species as the outcome of self-induced flavor conversions. If this occurs in the deepest stellar regions, all further matter effects occurring at larger distances get washed out. Moreover, such flavor equilibration is expected to have a significant impact on the dynamics of the SN explosion, altering neutrino energy deposition behind the shock wave. A direct measurement of a SN neutrino burst will be extremely useful to shed light on the important question: Do neutrinos experience self-induced conversions?

The goal of our work is to propose a {model-independent} test to distinguish between the pure matter effect scenario versus a complete flavor equalization during the accretion phase. We do this in a way that is agnostic to the fitting formulae for the neutrino spectra. We consider three kinds of detectors proposed for low-energy neutrino physics and astrophysics, viz., a water Cherenkov  (WC) detector with fiducial mass of 374 kton, e.g., the Japanese project Hyper-Kamiokande~\cite{Abe:2011ts}, a  liquid scintillator detector (SC) with a mass of 20 kton, e.g., the Chinese project JUNO~\cite{An:2015jdp}, and a Liquid Argon Time Projection Chamber (LAr TPC) with a mass of 40 kton, like the project DUNE in the United States~\cite{Acciarri:2015uup}. The neutrino-proton elastic scattering (pES) events at the scintillation detector, i.e., $\nu+p \to \nu+p$, is a neutral current process that is unaffected by flavor conversions and can be used as a measurement of the total neutrino fluence. We compare this to the events in the charged current channels, i.e., the inverse beta decay (IBD) process $\bar\nu_e + p \to n + e^+$, which is the dominant detection process for $\bar\nu_e$ in a water Cherenkov  detector, and with $\nu_e + {}^{40}Ar \to {}^{40}K^{\ast} + e^-$ charged-current process (ArCC) which is the dominant detection channel for $\nu_e$ in LAr detector.  We show that the ratio of events in these different detection channels can distinguish between complete flavor equalization and a pure matter effect driven flavor conversion, as long as spectral differences among the different species are not too small.

This paper is organized as follows. In Sec.\,\ref{sec:simul} we introduce our numerical benchmark models for SN neutrino emission, based on state-of-the-art SN simulations. In Sec.\,\ref{sec:osciscen} we present the expectations for the flavor conversions in the presence of matter effects or in the case of complete flavor equilibration. In Sec.\,\ref{sec:detection} we review the main features and detection channels for the three classes of large underground detectors we will consider for our analysis. In Sec.\,\ref{sec:recon} we review the reconstruction of  neutrino fluxes from the observed events in the three detectors. The experts can directly skip to Sec.\,\ref{sec:identflav}, where we present our main result, a model-independent technique to discriminate between the two oscillation scenarios, using a ratio of events from different detection channels. Finally, in Sec.\,\ref{sec:summary} we summarize our results and conclude.

\begin{figure}[!t]
\begin{centering}
\includegraphics[width=0.9\textwidth]{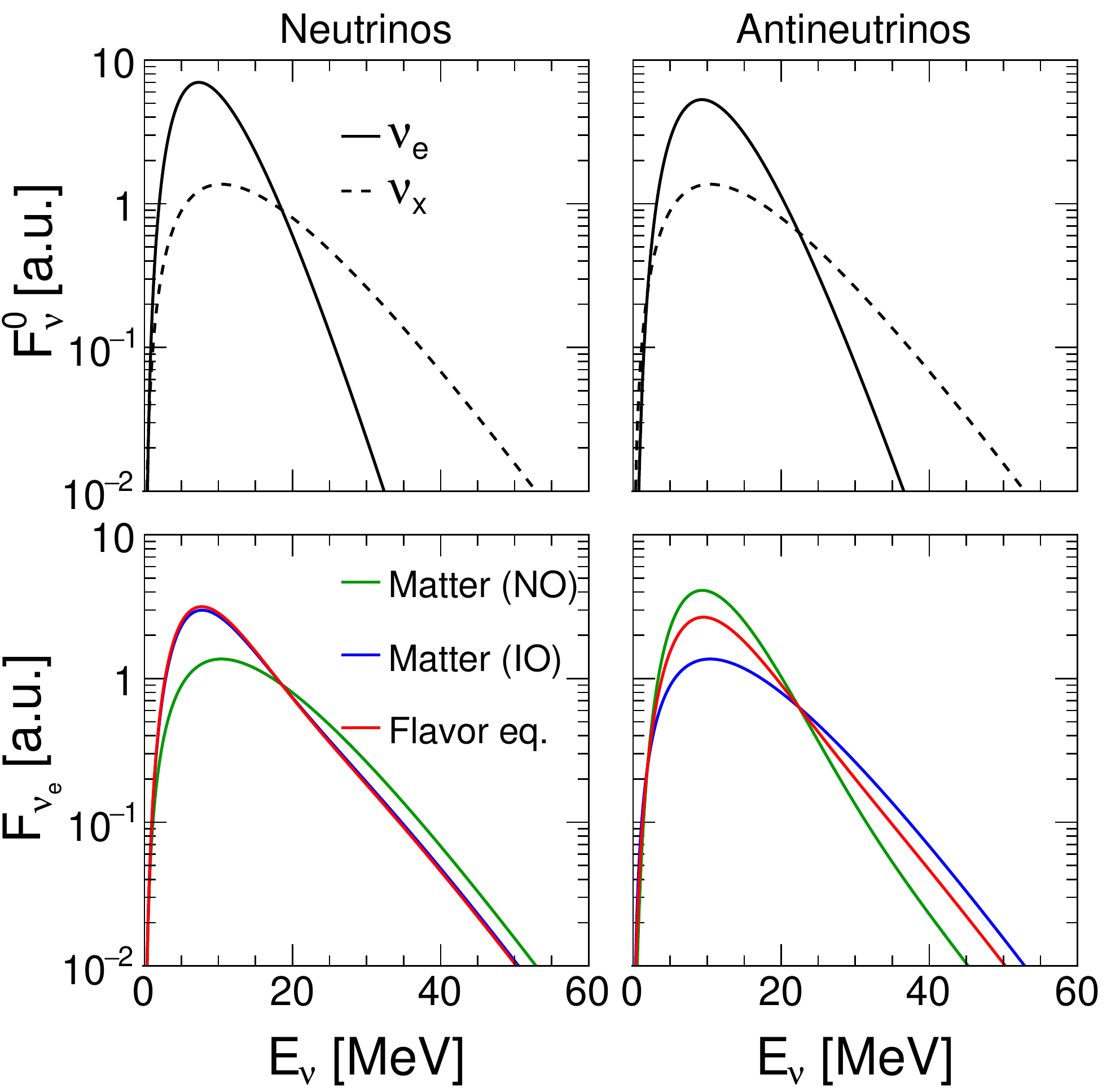}
\end{centering}
\caption{Wroclaw model: Neutrino fluence spectra for the different flavors,  unoscillated (top panels) and for different oscillation scenarios (bottom panels).} 
\label{fig:spectrum_basel}
\end{figure}  
\begin{figure}[h]
\begin{centering}
\includegraphics[width=0.9\textwidth]{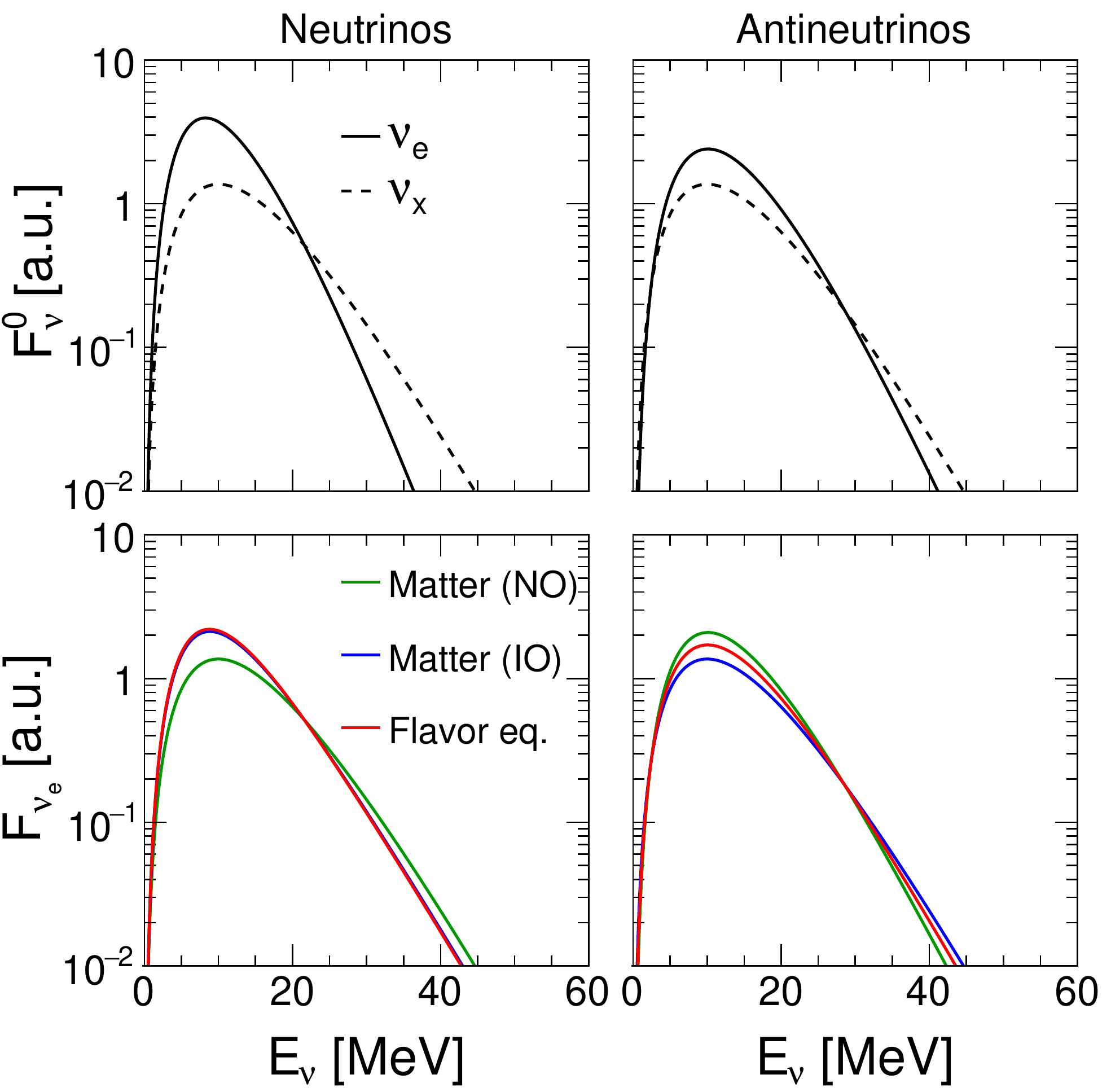}
\end{centering}
\caption{Garching model: Neutrino fluence spectra for the different flavors, unoscillated (top panels) and for different oscillation scenarios (bottom panels).} 
\label{fig:spectrum_garching}
\end{figure}

\section{Supernova Neutrino Fluences}
\label{sec:simul}

We consider the unoscillated time-integrated neutrino flux spectra emitted by the SN, i.e., the energy-dependent fluences,
\begin{equation}
F^0_\nu (E) =\Phi_\nu^0\, f_\nu^0(E) \,\ ,
\end{equation}
where $\Phi_{\nu}^0$ is the  time- and energy-integrated number flux for an interval of time $\Delta t$ under consideration and the function $f^0_\nu(E)$ is the time-averaged energy spectrum in the same epoch, normalized such that $\int dE\ f^0_\nu(E) = 1$. Note that $\nu$ refers to the flavors $\nu_e,\,\nu_\mu,\,\nu_\tau,$ and their antiparticles, where we will approximate that non-electron flavor neutrinos and antineutrinos have identical fluences, i.e., $F^0_{\nu_{\mu,\tau}}=F^0_{\bar{\nu}_{\mu,\tau}}=F^0_{\nu_x}$. 

The main strength of the method we will introduce in this paper is that we do not need a model or fit for the energy spectra $F^0_\nu(E)$. Usually, one relies on a so-called $\eta$-fit or an $\alpha$-fit to $f^0_\nu(E)$, in terms of distorted Fermi-Dirac or Maxwell-Boltzmann distributions, respectively~\cite{Keil:2002in}. Then one attempts to reconstruct the flavor mixing scenario by performing a joint-fit for the oscillation parameters and the parameters contained in these functions. This procedure is useful if the energy spectra predicted by the simulations are accurate representations of real SNe. However, such methods still fail to quantify what would happen if the simulations, though they may be extremely precise~\cite{Tamborra:2012ac}, are not accurate in their prediction of $F_\nu^0(E)$. Our paper is an attempt to address this issue and show that one can reconstruct the flavor mixing scenario using SN neutrino data, without assuming a functional form for $F_\nu^0(E)$.

\begin{table}[!h]
 \caption{Spectral-fit parameters for the neutrino and antineutrino fluxes integrated over the accretion phase of simulation of the of a $18 M_{\odot}$ star by the Wroclaw supernova project (W) and a $15 M_{\odot}$ star by the Garching group (G).}
\label{Tabsim}
\begin{center}
\begin{tabular}{lllllll}
\hline
Model& $\langle E_{\nu_e} \rangle$  (MeV) &  $\langle E_{\nu_x} \rangle$  (MeV) & $\Phi_{\nu_e} (\times 10^{56})$ & $\Phi_{\nu_x} (\times 10^{56})$ &
$\alpha_{\nu_e}$ & $\alpha_{\nu_x}$  \\
\hline
\hline
 W & 9.5 &15.6& 8.53& 3.13& 3.4& 2.0  \\
 G & 10.9 & 14.0 & 5.68  & 2.67 & 3.1 & 2.5 \\
 \hline
\end{tabular}
$ $ 
$ $ 
$ $

$ $ 

$ $ 

\begin{tabular}{lllllll}
\hline
Model& $\langle E_{\bar\nu_e} \rangle$  (MeV) &  $\langle E_{\bar\nu_x} \rangle$  (MeV) & $\Phi_{\bar\nu_e} (\times 10^{56})$ & $\Phi_{\bar\nu_x} (\times 10^{56})$ &
$\alpha_{\bar\nu_e}$ & $\alpha_{\bar\nu_x}$  \\
\hline
\hline
W & 11.6 &15.6& 7.51& 3.13& 4.0& 2.0  \\
G & 13.2 & 14.0 & 4.11  & 2.67 & 3.3 & 2.5 \\
 \hline
\end{tabular}
\label{tab:fluxes}
\end{center} 
\end{table}

As we have emphasized, the fitting parameters to the spectra are not needed for our analysis. However, to generate the pseudo-data that we use to illustrate our proposed technique, we adopt as test-cases two long-term SN simulations performed by the Wroclaw supernova project (W)~\cite{Fischer:2009af} and by Garching group (G)~\cite{Serpico:2011ir}, respectively.  The unoscillated (anti)neutrino  energy spectra,  $F^0_{\nu}$,  
are shown in the upper panels of  Fig.~\ref{fig:spectrum_basel} for the W model, and of Fig.~\ref{fig:spectrum_garching} for the G model, respectively.
Left panels refer to neutrinos, while right panels to antineutrinos.  The key difference between the two models is that the W model exhibits larger differences in the spectral features of the different species, while G model shows more similar neutrino spectra. For these simulations, the time-averaged energy spectrum over the accretion phase is parametrized as in~\cite{Keil:2002in}
\begin{eqnarray}
f^0_\nu(E) = \frac{1}{\langle E_\nu\rangle}\frac{(1+\alpha_\nu)^{1+\alpha_\nu}}{\Gamma(1+\alpha_\nu)}\left(\frac{E}{\langle E_\nu\rangle}\right)^{\alpha_\nu} \nonumber 
\exp\left[-(1+\alpha_\nu)\frac{E}{\langle
E_\nu \rangle}\right]\, ,
 \label{eq:varphi}
\end{eqnarray}
where $\langle E_\nu \rangle$ is the neutrino average energy, and the pinching parameter $\alpha_\nu $ is~\cite{Keil:2002in}
\begin{equation}
\alpha_\nu=\frac{2\langle E_\nu \rangle^2-\langle E_\nu^2\rangle}{\langle E_\nu^2\rangle-
\langle E_\nu\rangle^2} \, . \label{alphadef}
\end{equation} 
The values of the parameters are listed in Table~\ref{Tabsim}. Note that our choice of models is extremely conservative, in that even our optimistic model W has smaller differences between flavors, e.g., average energies of $\langle E_{\nu_e}\rangle=9.5$\,MeV, $\langle E_{\bar{\nu}_e}\rangle=11.6$\,MeV,  and $\langle E_{\nu_x}\rangle=15.6$\,MeV, than what was previously considered.

\section{Oscillation Scenarios: Matter Effect or Flavor Equalization}
\label{sec:osciscen}

In order to characterize the SN neutrino flavor conversions we assume a standard $3 \nu$ framework. Wherever needed we use the neutrino mixing parameters given by the recent global analysis~\cite{Capozzi:2018ubv}. In particular, because the ordering of the neutrino masses is still unknown, we will consider both the cases of normal ordering (NO) where $\Delta m^2_{\rm atm} = m_3^2 - m_{1,2}^2>0$, and the case of inverted mass ordering (IO) where   $\Delta m^2_{\rm atm}<0$. In the following, we will assume that at the time of the next galactic SN explosion,  the neutrino mass ordering will be known. This is a reasonable assumption given that the future supernova neutrino detection itself would give an independent indication of the neutrino mass ordering by exploiting the neutronization burst (see~\cite{Kachelriess:2004ds}). Of course, there are also several terrestrial experiments that aim to measure this quantity with a horizon of 10 years.
 
The flavor composition of neutrinos emitted by the SN are changed by self-induced and matter effects during neutrino propagation. The self-induced effects take place within $r \sim \mathcal{O}(10^{2})$~km from the neutrinosphere, whereas the MSW transitions take place at larger radii at $r \sim 10^{4}$--$10^{5}$~km.  As the self-induced and matter effects are widely separated in space, they can be considered
independently of each other. In general, one can write the oscillated neutrino fluences as ~\cite{Dighe:1999bi} 
\begin{equation}
F_{\nu_e} = P_{ee} F^{0}_{\nu_e} + (1-P_{ee}) F^0_{\nu_x} \,\ ,
\end{equation}
in terms of the original neutrino fluences $F^0_\nu$ and of the survival probability of 
electron neutrinos $P_{ee}$.  An analogous expression also holds for antineutrinos in terms of 
the electron antineutrino survival probability ${\bar P}_{ee}$,
\begin{equation}
F_{\bar{\nu}_e} = \bar{P}_{ee} F^{0}_{\bar{\nu}_e} + (1-\bar{P}_{ee}) F^0_{\nu_x} \,\ .
\end{equation}
These probabilities encode both self-induced and matter effects in the SN. 

During the accretion phase, we have two possible limiting cases.
If self-induced flavor conversions are suppressed by the multi-angle effects associated with the dense ordinary matter~\cite{EstebanPretel:2008ni},
the SN neutrino fluxes will be processed by only the matter effects while passing through the outer layers of the star. In this case, that we will denote by ME, the survival probabilities depend on the $1$-$2$ mixing angle, $\sin^2 \theta_{12}\simeq 0.3$, and on the neutrino mass ordering. In this case, the values of the survival probability for $\nu_e$ and $\bar\nu_e$ for the two mass orderings are reported in Table~\ref{tab:survival}. For simplicity, here we neglect additional matter
effects occurring if SN neutrinos cross the Earth before their detection. In fact, for the models described in Sec.\,\ref{sec:simul}, these Earth matter effects have a small impact on the oscillation pattern~\cite{Borriello:2012zc}. 
On the other hand, if fast flavor conversions or temporal instabilities act in the deepest SN regions unimpeded by matter effects therein, they tend to equalize the different fluences, i.e.,
 \begin{equation}
 F^0_{\nu_e,\,\nu_\mu,\,\nu_\tau} \to F_{\nu_e}=F_{\nu_\mu}= F_{\nu_\tau} = \frac{F^0_{\nu_e}+ F^0_{\nu_\mu}+ F^0_{\nu_\tau}}{3} \,\ ,
 \end{equation}
corresponding to $P_{ee} = 1/3$, independent of the mass ordering. Strictly speaking,  the flavor equalization cannot  occur simultaneously for $\nu$ and $\bar\nu$, so that if ${\bar P}_{ee} = 1/3$ then ${P}_{ee} > 1/3$ due to the lepton number conservation in self-induced conversions~\cite{Hannestad:2006nj}. 
To take account this possibility in the following analysis, we will account also for deviation with respect to the complete flavor equilibrium.
However, for simplicity, we take as working hypothesis that there is complete equalization in both channels. This scenario, dubbed as self-induced flavor equalization, is shown in Table \ref{tab:survival} labelled by FE. In the presence of partial flavor equalization, the total survival probability would be intermediate between the one given by only matter effect and the one obtained for complete flavor equalization. Our present analysis can be easily modified to test these other oscillation scenarios.
 
\begin{table}[!t]
 \caption{Survival probabilities for electron neutrinos, $P_{ee}$, and antineutrinos,
  $\bar{P}_{ee}$, in various two mixing scenarios: ordinary matter effects only (ME) and self-induced flavor equalization (FE).}
\begin{center}
\begin{tabular}{llll}
\hline
Scenario&Mass Ordering&  $P_{ee}$ & $\bar{P}_{ee}$ \\
\hline
\hline
 ME & NO  & 0 & $\cos^2\theta_{12}\simeq0.7$ \\
 ME & IO & $\sin^2\theta_{12}\simeq0.3$ & 0 \\
 FE & either & 1/3 $\simeq$ 0.33 & 1/3 $\simeq$ 0.33\\
 \hline
\end{tabular}
\label{tab:survival}
\end{center} 
\end{table} 
 
Our analysis of matter effects here is based on considering adiabatic MSW resonances, as is appropriate for the accretion phase. In general, however, more complicated intermediate scenarios due to nonadiabaticity, multiple resonances, and turbulence are possible (see, e.g., Ref.~\cite{Mirizzi:2015eza}). One would have to suitably generalize the expression for $P_{ee}$ ($\bar{P}_{ee}$) in the NO (IO), to span the allowed range and account for these possibilities, e.g., in the cooling phase.

A robust feature, across all SN simulations, is that at sufficiently large energies the non-electron (anti)neutrino fluence, $F^0_{\nu_x}$, dominates over the electron (anti)neutrino fluence and an inversion of this hierarchy is a model-independent signature of flavor conversion. One can define a  critical energy, $E_c$, such that in the high-energy tail, at $E> E_c$, one has the $F^0_{\nu_e} \ll F^0_{\nu_x}$.  In the antineutrino sector, ${\bar E}_c$ denotes the analogous critical energy. An inversion of this expected hierarchy of the fluences, e.g., $F^0_{\nu_e} \gg F^0_{\nu_x}$ instead of $F^0_{\nu_x} \gg F^0_{\nu_e}$, is a signature that flavor conversion must have occurred. Our proposed method will use this signature. This feature stems from the fact that non-electron (anti)neutrinos have fewer interactions and decouple deeper in the star which allow more of these non-electron neutrinos to have higher energies.

The precise value of $E_c$ or $\bar{E}_c$ is model-dependent. From  Fig.~\ref{fig:spectrum_basel} one sees that in the W model $E_c \simeq  {\bar E}_c \simeq 20$~MeV. Above the critical energies the differences between $F^0_{\nu_e}$ and $F^0_{\nu_x}$ are sizeable, being  $F^0_{\nu_e}/ F^0_{\nu_x} \simeq 0.009$ and $F^0_{\bar\nu_e}/ F^0_{\bar\nu_x} \simeq  0.048$ at $E \simeq 40$ MeV. These large flux differences would make it easier to detect oscillation effects. For the G model in Fig.~\ref{fig:spectrum_garching}, the critical energies are higher, being $E_c \simeq 25$ MeV and ${\bar E}_c \simeq 30$ MeV. Also, above these critical energies, one has $F^0_{\nu_e}/ F^0_{\nu_x} \simeq 0.142$ and $F^0_{\bar\nu_e}/ F^0_{\bar\nu_x} \simeq 0.549$ at $E \simeq 40$. Consequently, oscillation effects will be harder to observe for the G model. Nevertheless, with sufficient statistics in the highest energy bins one may be able to extract oscillation-dependent information even in this case.

From Figs.~\ref{fig:spectrum_basel} and \ref{fig:spectrum_garching} (lower panels) and Table~\ref{tab:survival} we notice that for NO there are differences in the survival probabilities for both $\nu$ and $\bar\nu$, so that one can exploit both neutrino and antineutrino data to disentangle the different oscillation scenarios. Unfortunately for an IO,  $P_{ee}=0.3$ for the matter effects scenario while $P_{ee}={0.33}$ for flavor equalization. Therefore, these two oscillation scenarios would be practically indistinguishable using $\nu$ data alone. In this case one should rely on the $\bar\nu$ data, where  ${\bar P}_{ee}= 0$ for the matter effects scenario.

\section{Neutrino detection}  
\label{sec:detection}

In this section we describe the main aspects and ingredients of 
our calculations  of supernova neutrino event rates.
The oscillated SN neutrino fluxes at the Earth, $F_\nu$,  
must be convolved with the differential cross section   
$\sigma$ of the neutrino interaction process, 
as well as with the energy resolution function $W$ of the detector, 
and the efficiency $\varepsilon$ in order to compute the 
observable event rates~\cite{Fogli:2004ff}: 
\begin{equation}  
N_{\rm ev} = F_\nu \otimes \sigma \otimes W \otimes \varepsilon\ .  
\label{Conv}  
\end{equation}  

We will now describe the main 
characteristics of the three types of detectors we have used to calculate the signals. 
For each detector we will describe only the detection channel we will use for our analysis. In particular,
we will assume $\varepsilon=1$ above the energy threshold for the considered channel. Moreover, we  assume that other 
neutrino interaction channels can be separated at least on a statistical basis.
We will take $d=10$ kpc as the benchmark distance to a galactic SN, unless otherwise stated. 

Hereafter we consider only statistical errors, neglecting systematics connected to, e.g., cross section or energy scale uncertainties. In particular, the cross sections for pES or ArCC are known with an error $\gtrsim 10\%$, which, as shown in the last part of this paper, is comparable to the statistical uncertainties of events ratios. We assume that dedicated calculations or direct measurements will be performed in the future in order to improve current knowledge of cross sections. Such developments are needed not only for the purposes of this work, but more generally for getting the most out of the next supernova neutrino burst.

\subsection{Scintillation detector}
In this work we consider a  liquid scintillation detector
with a fiducial mass of 20 kton~\cite{An:2015jdp} and a Gaussian energy resolution as function of the visible energy 
$E_{\rm vis}$, with
a width $\Delta$
\begin{equation}
\Delta_{\rm{SC}}/\textrm{MeV} = 0.03\sqrt{{E_{\rm vis}}/\textrm{MeV}} \,,
\label{sc-resolution}
\end{equation}
as proposed for the Chinese JUNO project~\cite{An:2015jdp}.
For our analysis it will be crucial to use 
 the elastic scattering of neutrinos on protons (pES)
\begin{equation}
\nu + p \to \nu + p \,\ .
\end{equation}

The contributions from all flavors and from both neutrinos and antineutrinos partly compensate the factor of four difference between the pES and IBD cross sections (see later). For this channel a low energy threshold is required, since the proton recoil energy $T_p \leq 2 E^2_{\nu}/m_p$ is suppressed by the nucleon mass. Additionally, a precise measurement of the proton quenching factor is mandatory for a good energy reconstruction. This channel originally proposed in Ref.~\cite{Beacom:2002hs}, becomes especially important in the context of disentangling self-induced oscillations, as emphasized in Ref.~\cite{Dasgupta:2011wg}. 

The observed event spectrum for the neutral current pES channel is given by
\begin{equation}
\frac{dN_{\rm pES}}{dE_{\text{vis}}} = N_p\int_{0}^{+\infty}dT'_p \frac{dT_p}{dT'_p}W(T'_p,E_{\text{vis}})\int_{E_{\nu}^0}^\infty dE_{\nu} F_{\rm pES}(E_\nu)
\frac{d\sigma_{\text{pES}}(E_{\nu},\,T_p)}{dT_p}\,,
\label{pES_spectrum}
\end{equation}
where, independent of the flavor conversion scenario (as long as one doesn't invoke exotic physics, e.g., sterile neutrinos), one has
\begin{equation}
F_{\rm pES} \equiv  4 F^0_{\nu_x} +
F^0_{\bar\nu_e} + F^0_{\nu_e} \,\ ,
\label{eq:fpes}
\end{equation}
and $N_p$ = $1.44\times 10^{33}$ is the number of free protons in the detector, $T_p$ is the true proton kinetic energy, while $T'_p$ is quenched proton kinetic energy, $E_{\text{vis}}$ is the observed quenched proton kinetic energy, with $E^0_\nu=\sqrt{T_pm_p/2}$ being the minimum neutrino energy to produce a proton with a recoil energy $T_p$, and $\sigma_{\rm pES}$ is the cross section~\cite{Dasgupta:2011wg}, with $W(T'_p,E_{\text{vis}})$ being the energy resolution function.
For this channel we consider a lower energy threshold of $E_{\rm vis}\geq0.2$ MeV due to irreducible backgrounds~\cite{Dasgupta:2011wg}.

\begin{figure}[!t]
\includegraphics[width=\textwidth]{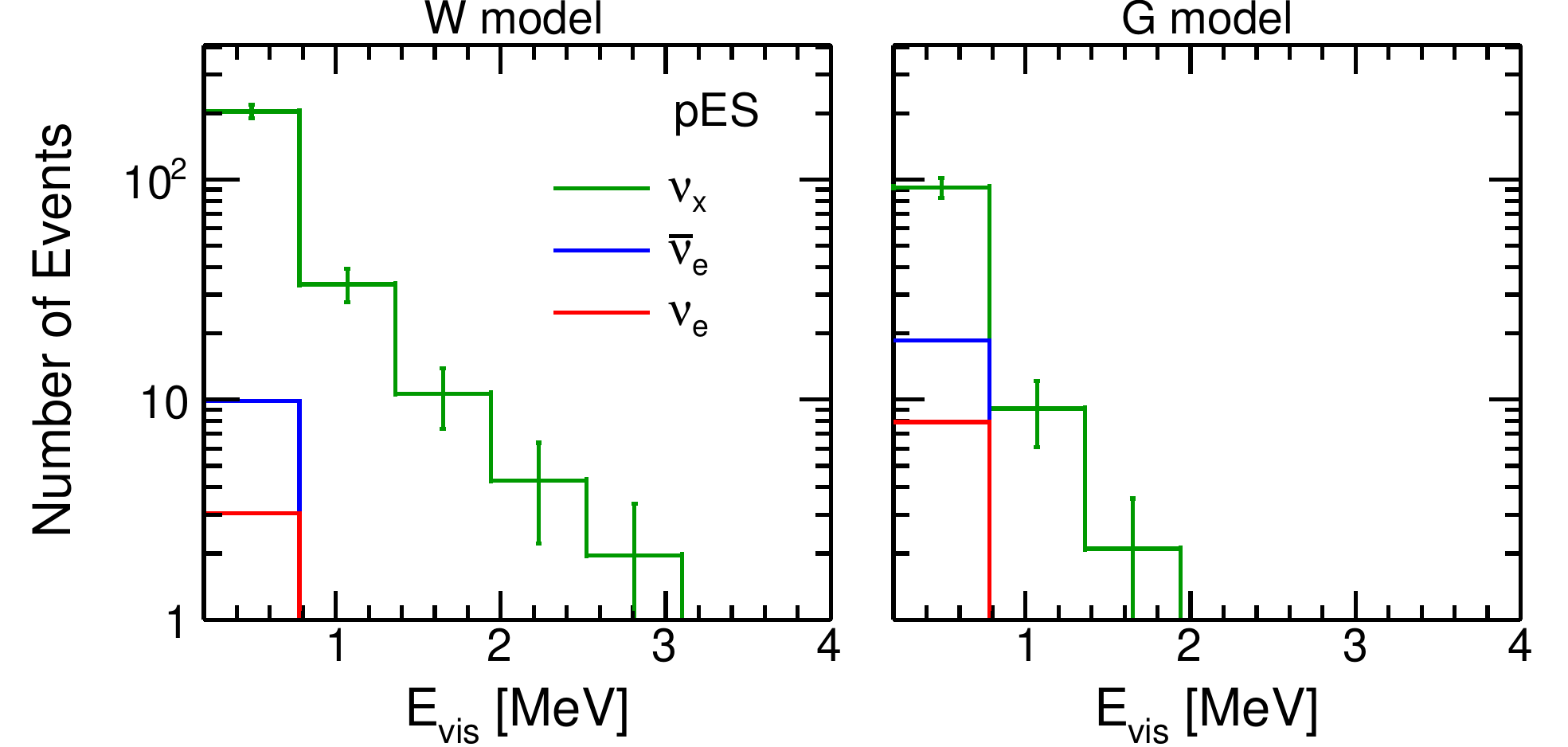}
\caption{Observed spectrum from proton elastic scattering in JUNO, for Wroclaw (left) and Garching (right) model.
The red curve represents the contribution of $\nu_e$, the blue curve is for $\bar\nu_e$ and the green for $\nu_x$.} 
\label{fig:observed_events_pES}
\end{figure}

Figure~\ref{fig:observed_events_pES} shows the events distribution for pES events in a scintillation detector for  W (left panel) and G model (right panel). The total number of events for the former (latter) model is 270 (140).
 We show the contribution to the total number of events from $\nu_e$, $\bar{\nu}_e$ and $\nu_x$ separately. As one can see, the signal in high-energy tail is dominated by the $\nu_x$ contribution. However, contrary to previous works based on more energetic $\nu_x$ spectra~\cite{Dasgupta:2011wg}, the relative weight of electron flavors is not completely negligible, especially for the G model. Therefore, in the following we will include all the neutrino species in the evaluation of events rate for pES. 

Scintillation detectors, are also sensitive to other detection channels, such as the IBD events due to $\bar{\nu}_e$ on protons, which are also the dominant ones. Nevertheless pES events can be easily distinguished from other channels as shown in \cite{Dasgupta:2011wg}. Since water Cherenkov detectors have  much larger statistics for IBD, we will show the observed IBD event distribution for this latter type of detector, as described in the next section.

\begin{figure}[h]
\includegraphics[width=\textwidth]{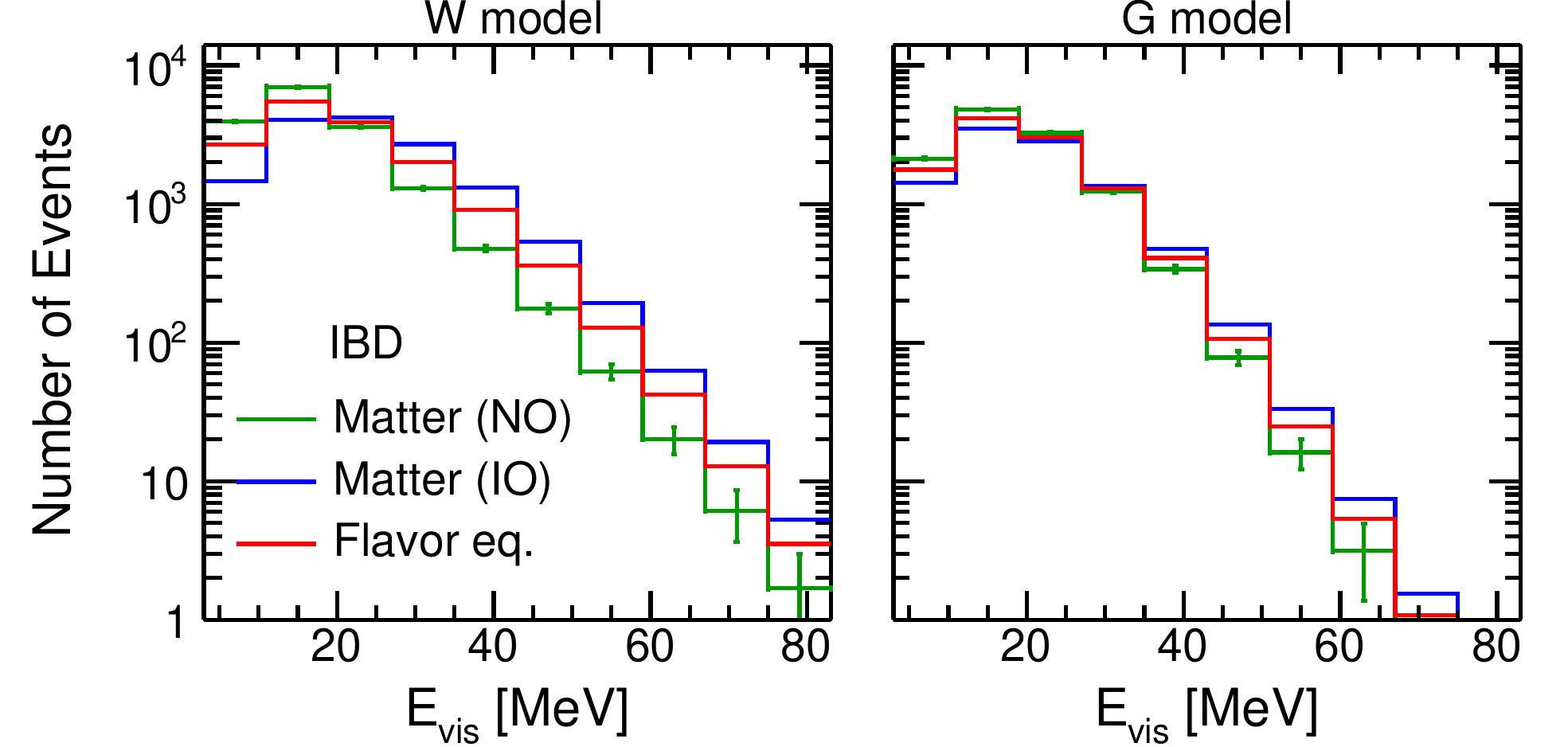}
\caption{Observed spectrum from inverse beta decay in Hyper-Kamiokande for Wroclaw (left) and Garching (right) model. The green (blue) curve represents matter effects only in NO (IO), while the red curve refers to complete flavor equalization.} 
\label{fig:observed_events_IBD}
\end{figure}

\subsection{Water Cherenkov detectors}
In this work we consider a water Cherenkov detector \citep{Abe:2011ts,HyperK_new} with  an energy-resolution of 20\% for a
positron energy of 10~MeV~\cite{Abe:2011ts}, which corresponds to
\begin{equation}
\Delta_{\rm{WC}}/\textrm{MeV} = 0.6\sqrt{E_{\rm vis}/\textrm{MeV}} \,\ ,
\end{equation}
where $E_{\rm vis}$ is the true positron energy,
and a total fiducial mass of 374 kton, as proposed 
for the Japanese Hyper-Kamiokande project~\cite{Abe:2011ts}.
Correspondingly, the number of free protons in the detectors
is $N_p= 2.48 \times 10^{34}$. The dominant channel for supernova 
neutrino detection is
the IBD of ${\bar\nu}_e$'s
\begin{equation}
{\bar\nu}_e + p \to n+ e^+ \,\ ,
\end{equation}
where the positrons are detected through 
photons produced in the scintillation material.
The observed event spectrum for the IBD channel is given by
\begin{equation}
\frac{dN_{\rm IBD}}{dE_{\text{vis}}} = N_p\int_{E_T}^\infty dE_{\nu} 
F_{\text{IBD}}(E_{\nu})\sigma_{\text{IBD}}(E_{\nu})\;
W(E_\nu-0.782 \,\ \textrm{MeV},\,E_\mathrm{vis}),
\label{IBD_spectrum}
\end{equation}
where $\sigma(E_{\nu})$ is differential the cross section as calculated in \cite{Strumia:2003zx} and $E_{\text{vis}}$ is the energy of scintillation photons (visible energy). Here we are neglecting nucleon recoil, i.e., we are assuming that the positron energy is equal to $E_{\nu}-0.782 \,\ \textrm{MeV}$. $W$ is the energy resolution function that implements a Gaussian smearing of the visible energy with a width $\Delta_{\rm WC}$.

Furthermore, since in this work we are interested in the high energy tail of neutrino spectrum ($E_\nu\gtrsim 25$ MeV), we neglect any background contribution for this channel.
The antineutrino flux $F_{\rm IBD}$ probed by this detection channel
depends on the oscillation scenario (see Table~\ref{tab:survival}), namely
\begin{equation}
F_{\rm IBD} \equiv
 \left\{
\begin{array}{ll}
 0.7 F^0_{\bar\nu_e} + 0.3 F^0_{\nu_x}  & \textrm{matter effects only, with NO} \\
F^0_{\nu_x} & \textrm{matter effects only, with IO}\\
0.33 F^0_{\bar\nu_e} + 0.66 F^0_{\nu_x} & \textrm{flavor eq.} \\
\end{array}\right.
 \,.
\label{eq:fibd}
\end{equation}

Figure~\ref{fig:observed_events_IBD} shows the observed distribution of events in a water Cherenkov detector for three different oscillation scenarios and for both W
(left panel) and G (right panel) models. 
Assuming flavor equalization, the total number of events for the former (latter) model is 15500 (10800). We realize that due to the high-statistics of events, the different oscillation scenarios can be distinguishable by a spectral analysis alone. 
 
A final remark is in order. In this work we are assuming that the water inside the detector will be doped  with gadolinium, which will allow a 90\% tagging
efficiency for IBD events \cite{Beacom:2003nk}, making contamination from other channels negligible in terms of the performance of the method proposed herein.

\subsection{Liquid Argon Time Projection Chambers}  
In this work we consider a  LAr TPC detector with an energy resolution of~\cite{Amoruso:2003sw} 
\begin{equation}
\Delta_{\rm{LAr}}/\textrm{MeV} = 0.11\sqrt{{E_{\rm vis}}/\textrm{MeV}} + 
0.02\, E_{\rm vis} / {\rm MeV} \,\ ,
\label{LAr-resolution}
\end{equation}
and a fiducial volume of 40 kton, 
as assumed for the DUNE project in the United States~\mbox{\cite{Acciarri:2015uup,Strait:2016mof,Acciarri:2016ooe}}.
 Note that we are assuming the same energy resolution of ICARUS~\cite{Amoruso:2003sw}, since DUNE low energy capabilities are still under investigation. Given the challenges in reconstructing low electron energy in DUNE detector, the energy resolution might turn out to be worse than what expected in ICARUS. In such a case, the power of the method proposed in this paper will be affected accordingly. 

\begin{figure}[!t]
\includegraphics[width=\textwidth]{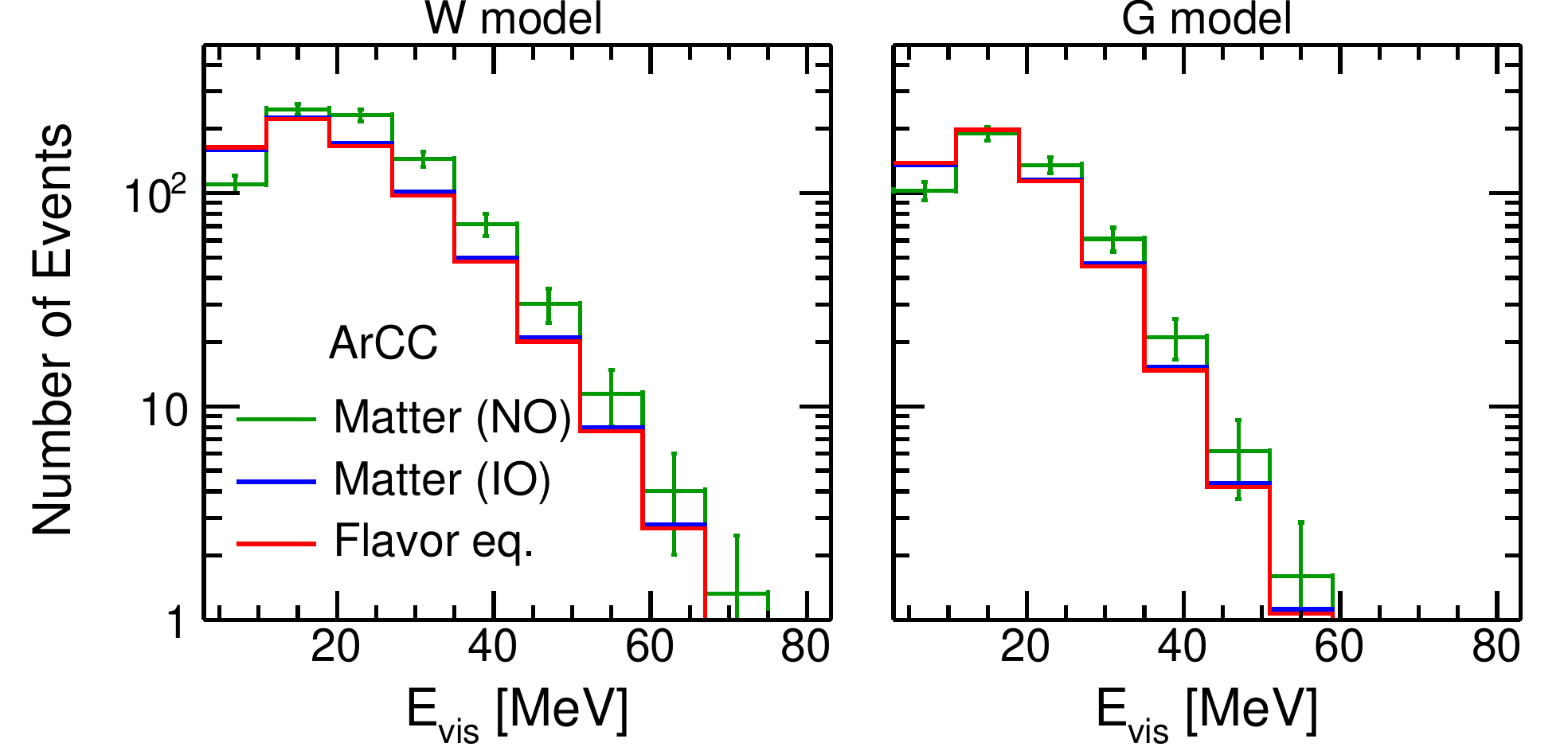}
\caption{Observed event spectrum from charged current on Argon in DUNE for Wroclaw (left) and Garching (right) model. The green (blue) curve represents matter effects only in NO (IO), while the red curve refers to complete flavor equalization.} 
\label{fig:observed_events_ArCC}
\end{figure}

LAr TPC detectors  are particularly sensitive to SN electron 
neutrinos through their charged current interactions with Ar nuclei (ArCC) 
\begin{equation}  
 \nu_e + {}^{40}\text{Ar} \to {}^{40}\text{K}^{\ast} + e^- \,\ ,  
\end{equation}  
which proceed via the creation of an excited state of $^{40}$K and its  
subsequent gamma decay. The reaction threshold to reach the ground state of $^{40}$K is $Q_{gs}=1.505$~MeV. The total Q-value for the $i$-th excited state of $^{40}$K is given by $Q_{i}=Q_{gs}+\Delta E_i$, where $\Delta E_{i}$ is the difference between the excited state energy and the energy of its ground state. 
The observed event spectrum for the ArCC channel is given by
\begin{equation}
\frac{dN_{\rm ArCC}}{dE_{\text{vis}}} = N_{\rm Ar}\sum_{i=1}^{N_{\rm ex}}\int_{0}^\infty dE_{\nu} 
F_{\text{ArCC}}(E_\nu)\sigma^i_{\text{ArCC}}(E_{\nu})\;
W(E_{\text{vis}},T_e),
\label{ArCC_spectrum}
\end{equation}
where $N_{\rm ex}$ is the number of excited states of $^{40}K$, $T_e$ is the electron energy, $E_{\text{vis}}$ is the reconstructed electron energy, $\sigma^k_{\rm ArCC}$ is the cross section for the $k-$th excited state and $N_{\rm Ar}=5.98\times 10^{32}$ is the number of $^{40}$Ar nuclei in the detector. The list of excited states and of the strength of each transition is taken from~\cite{Bhattacharya:2009zz}.  Here we neglect any information coming from the detection of gamma rays coming from the de-excitation of $^{40}$K, which can only improve the event reconstruction capability. Analogously to IBD, we neglect background contributions.
The neutrino flux combination probed the ArCC process in the different oscillation scenarios (see Table~\ref{tab:survival})
is given by
\begin{equation}
F_{\rm ArCC} \equiv
 \left\{
\begin{array}{ll}
 F^0_{\nu_x}  & \textrm{matter effects only,  with NO} \\
0.3 F^0_{\nu_e} + 0.7 F^0_{\nu_x} & \textrm{matter effects only, with IO}\\
0.33 F^0_{\nu_e} + 0.66 F^0_{\nu_x} & \textrm{flavor equalization} \\
\end{array}\right.
 \,.
\label{eq:farcc}
\end{equation}
 
In Fig.~\ref{fig:observed_events_ArCC} we show the events distribution for ArCC.
Assuming flavor equalization, the total number of events for W (G) model is 730 (520).
As expected from Eq.~(\ref{eq:farcc}), the event spectra expected for matter effects only (with IO) and flavor equalization are practically indistinguishable.

Although DUNE is sensitive to other channels, such as neutrino elastisc scattering on electrons or $\bar{\nu}_e$ charged current scattering on $^{40}$Ar, 80-90\% of the total are represented by $\nu_e$ events~\cite{Migenda:2018ljh}. We can thus safely neglect contamination from other channels. In particular, ArCC events are in principle distinguishable from elastic scattering on electrons if de-excitation gamma rays are taken into account.

\section{Reconstructing neutrino fluxes}  
\label{sec:recon}
  
From the observed event distributions, described in the previous section, we can reconstruct the oscillated neutrino fluences without any fitting. We briefly review how such a process is performed for the different detection channels.

\subsection{Proton Elastic Scattering}
Let us first start with the pES channel.
The true neutrino spectrum is given by $F_{\rm pES}$ in Eq.~(\ref{eq:fpes}). One must note that the much larger event rates from IBD in fact have to be tagged and subtracted to extract the pES event rate that we calculated above. We assume that this will be done. Our analysis can also be appropriately modified by considering that a known fraction of contaminant IBD events adds to the measured $F_{\rm pES}$; the relative weight depending on the tagging efficiency and the relative cross sections for IBD and pES. We do not pursue this issue here.

We denote  the one obtained through the reconstruction procedure as $\tilde{F}_{\rm pES}$ . 
 Let us assume that the observed spectrum of events is divided into $N$ bins and that the observed number of events in the $i-$th bin is $N_{pES}^i$. We define the extrema of the $i-$th bin as $[T'^i_{p},T'^{i+1}_{p}]$ and its midpoint as $\bar{T'}^i_p$. The corresponding extrema and midpoint for the true proton kinetic energy are  $[T^i_{p},T^{i+1}_{p}]$ and $\bar{T}^i_p$, respectively. Finally, we define the extrema and midpoint for the neutrino energy bins as $[E^i_{\nu},E^{i+1}_{\nu}]$ and $\bar{E}^i_\nu$, respectively, where $E^i_{\nu}=\sqrt{T^i_pm_p/2}$. Following the approach proposed in~\cite{Dasgupta:2011wg,Li:2017dbg}, we can calculate the neutrino fluence, i.e., the number of neutrinos per $\rm{cm}^{-2}\times \rm{MeV}^{-1}$, through the recursive equations:
\begin{eqnarray}
\left.\frac{d\tilde{F}_{\rm pES}}{dE_{\nu}}\right\vert_{\bar{E}_{\nu}^N}&=&\frac{N_{\rm pES}^N}{K_{NN}}\label{inversion_pES_1}\\
\left.\frac{d\tilde{F}_{\rm pES}}{dE_{\nu}}\right\vert_{\bar{E}_{\nu}^i}&=&\left(N_{\rm pES}^i+\sum_{j>i}\left.\frac{d\tilde{F}_{\rm pES}}{dE_{\nu}}\right\vert_{\bar{E}_{\nu}^j}K_{ij}\right)/K_{i,i}\,,
\label{inversion_pES_2}
\end{eqnarray}
where
\begin{equation}
K_{i,j}=N_p\Delta T'^i_p\left.\frac{dT_p}{dT'_p}\right\vert_{\bar{T}'^i_p}\left. \frac{d\sigma_{\rm{\rm pES}}(E_\nu,T_p)}{dT_p}\right\vert_{(\bar{T}'^i_p,\bar{E}_{\nu}^j)}\,.
\label{Kmatrix}
\end{equation}
The flux obtained through Eqs.~(\ref{inversion_pES_1}) and (\ref{inversion_pES_2}) is dominated by the  ${\nu}_x$ with 
a subleading contribution of  $\bar{\nu}_e$ and $\nu_e$. Note that we are neglecting the energy resolution in such calculations. However, given the accurate energy reconstruction in scintillation detectors, its impact on the inversion procedure is small. 
Furthermore, in order to avoid instabilities in the process of inversion, the size of the bins has to be chosen such as the number of observed events in each of them is comparable. The statistical errors in each bin of $E_\nu$ are assumed to be equal to the statistical fluctuation correspondent to the observed number of events in the same bin, i.e., $\sqrt{N^i_{\rm pES}}$. 

\begin{figure}[!t]
\includegraphics[width=\textwidth]{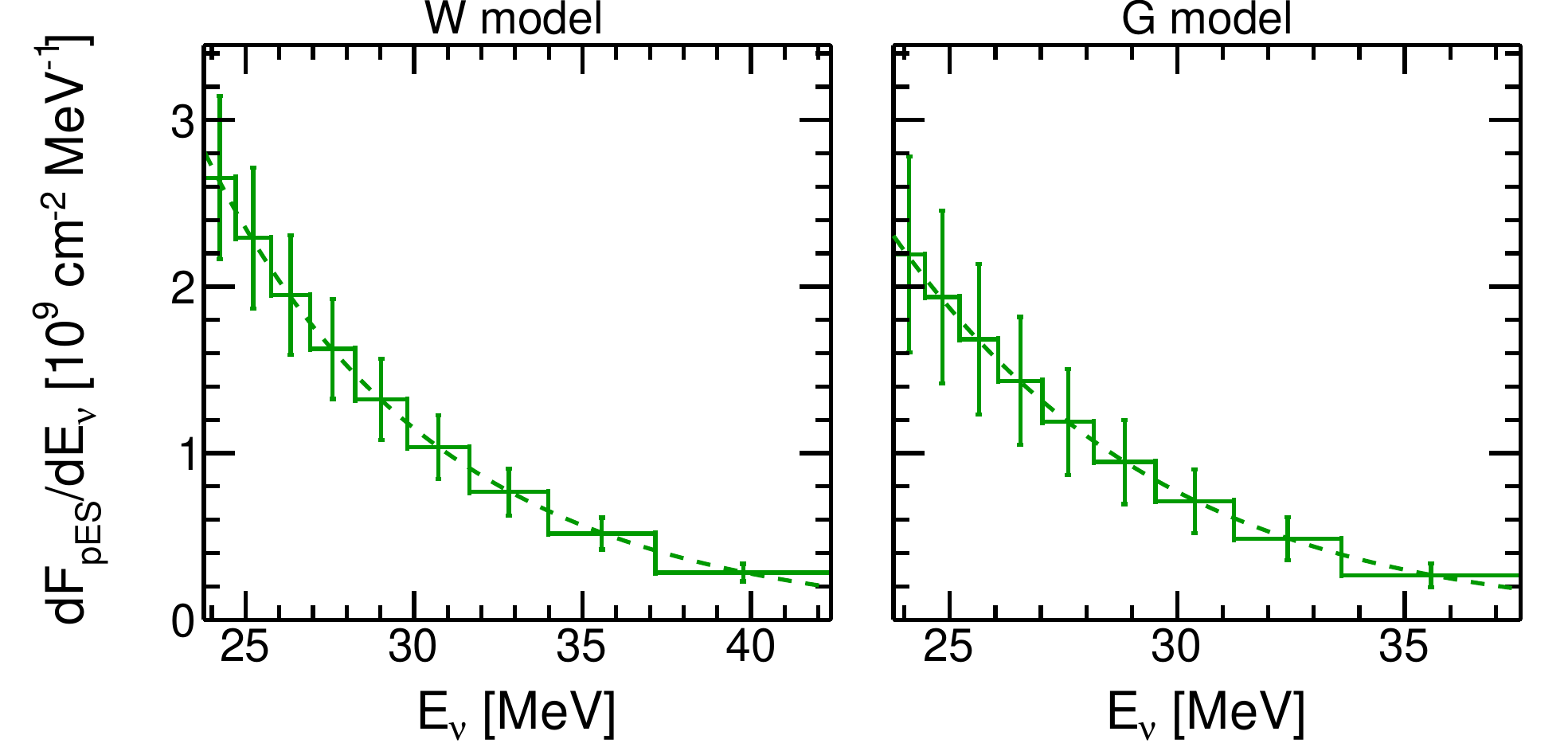}
\caption{Reconstructed neutrino energy spectrum ${\tilde F}_{\rm pES}$ from proton elastic scattering in JUNO for Wroclaw (left panel) and Garching (right panel) model. Dashed curves represent the true
neutrino flux $F_{\rm pES}$ in Eq.~(\ref{eq:fpes}).} 
\label{fig:pES_inverted}
\vspace{1.0cm}
\includegraphics[width=\textwidth]{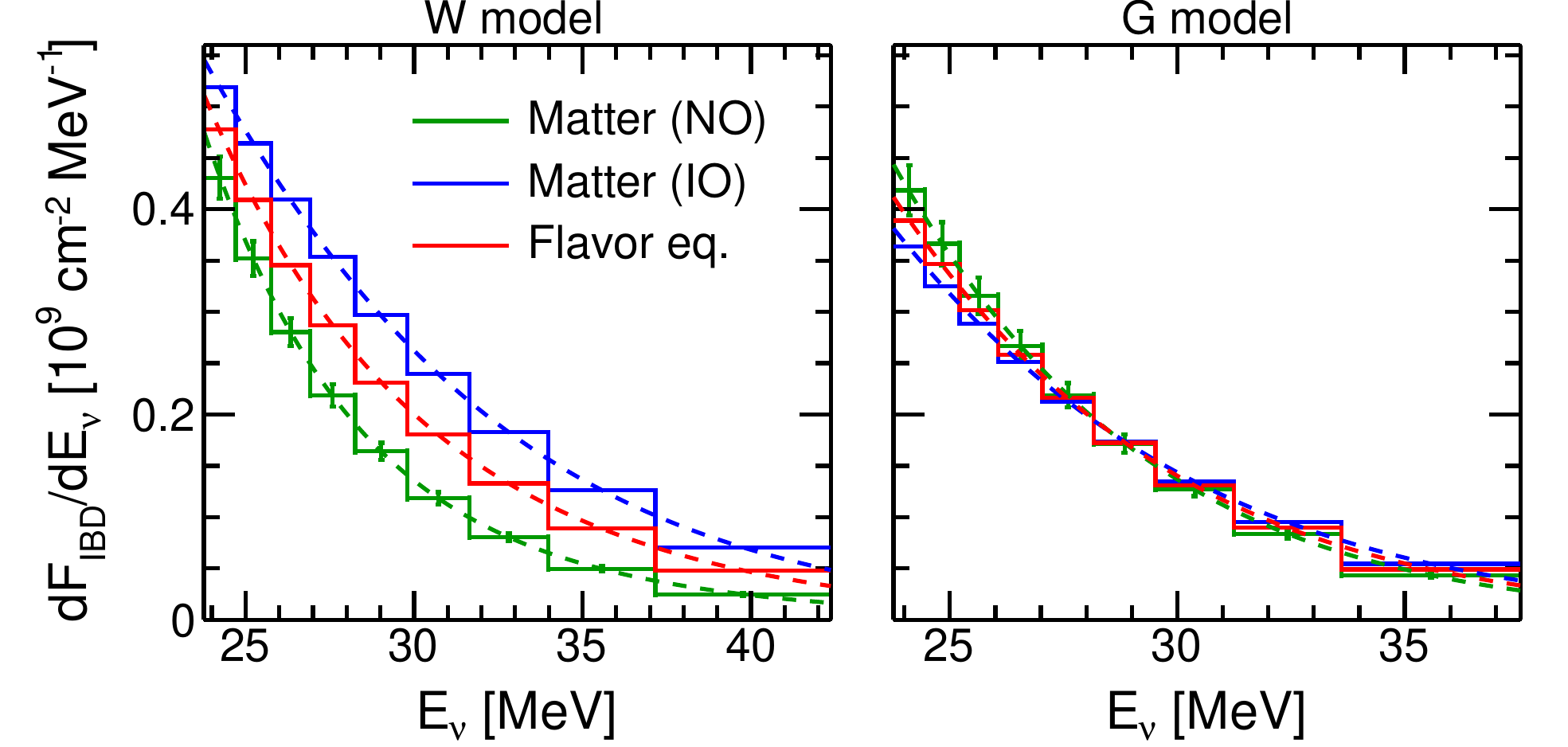}
\caption{Reconstructed neutrino energy spectrum ${\tilde F}_{\rm IBD}$ from inverse beta decay in Hyper-Kamiokande, assuming Wroclaw (left panel) and
Garching (right panel) models. The green (blue) curve represents matter effects only in NO (IO), while the red curve refers to complete flavor equalization. Dashed curves refer to the true neutrino flux
${F}_{\rm IBD}$ in Eq.~(\ref{eq:fibd}). } 
\label{fig:IBD_inverted}
\end{figure}  
\begin{figure}[!t]
\includegraphics[width=\textwidth]{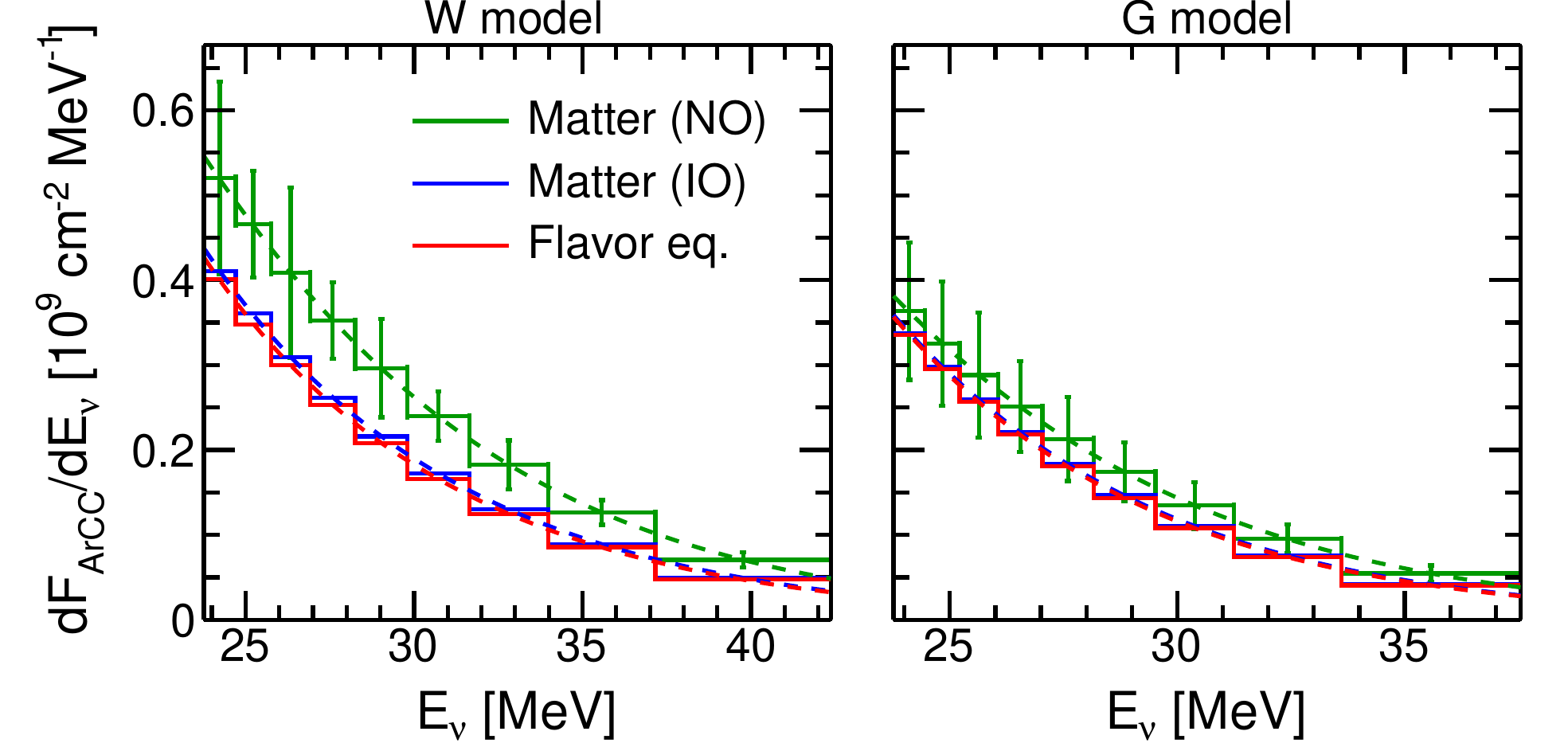}
\caption{Reconstructed neutrino energy spectrum ${\tilde F}_{\rm ArCC}$ from charged current on Argon in DUNE, assuming Wroclaw (left panel) and
Garching (right panel) models. The green (blue) curve represents matter effects only in NO (IO), while the red curve refers to complete flavor equalization. Dashed curves refer to the true neutrino flux
${F}_{\rm ArCC}$ in Eq.~(\ref{eq:farcc}).} 
\label{fig:ArCC_inverted}
\end{figure}

While the procedure outlined above seems straight-forward, several technical difficulties must be overcome. In Eqs.~(\ref{inversion_pES_1}) and (\ref{inversion_pES_2}), each
bin of neutrino energy receives a contribution from the fluence already
calculated at higher energy, whose statistical error must be added in quadrature.
This propagation of errors significantly increases the error especially in the low-energy bins.
However, it has been shown in~\cite{Li:2017dbg}
that a more sophisticated  unfolding procedure based on
singular value decomposition allows one
to get significantly smaller error-bars (especially in the low-energy bins) with respect to the ones
estimated by the simple propagation of error. Another issue concerns what one chooses as the higher end of the last energy bin. Choosing it to be equal to the highest observed neutrino energy or much too larger, both lead to biased reconstruction. There are also biases related to using the midpoint of the bin~\cite{Dasgupta:2011wg}. One may be able to reduce these biases by assuming some priors. These are technically important issues but our focus in this paper is not to make these improvements but rather to show what one can do if such a reconstruction exists. With this point of view, we will assume that on average the reconstruction is unbiased and errors are dominantly statistical. The reconstructed flux ${\tilde F}_{\rm pES}$ obtained with this procedure is shown in Fig.~\ref{fig:pES_inverted} for the W (left panel) and G (right panel) models. From the comparison with the true neutrino flux ${F}_{\rm pES}$ one realizes that, despite the caveats mentioned above, the reconstruction is quite accurate.

\subsection{Inverse Beta Decay}

We follow the approach proposed in~\cite{Li:2017dbg} to reconstruct $F_{\rm IBD}$ defined in Eq.~(\ref{eq:fibd}), 
obtaining the reconstructed flux in a water Cherenkov detector
\begin{equation}
\left.\frac{d{\tilde F}_{\rm IBD}}{dE_{\nu}}\right\vert_{\bar{E}_i}=\frac{1}{N_p\sigma^{\rm tot}_{\rm IBD}(\bar{E}_i)}\frac{N^i_{\rm IBD}}{\Delta E^i_{\rm vis}}\,,
\label{inversion_IBD}
\end{equation}
where we are neglecting the energy resolution of the detector. Note that we bin the events in order to obtain the neutrino flux for the same $E^i_\nu$ adopted in the case of pES. Therefore, we discard the low energy part of the supernova neutrino spectrum, to which the IBD is sensitive and whose statistical contribution is relatively strong. 
The reconstructed fluxes ${\tilde F}_{\rm IBD}$ and the original ones ${F}_{\rm IBD} $   are shown in Fig.~\ref{fig:IBD_inverted} for different oscillation scenarios
for W (left panel) and G (right panel) models. One can see that for the W model the reconstructed fluxes  show seizable differences in the different oscillations scenarios. Conversely, in the G case where spectral differences in the $\bar\nu$ channels are milder at high energies the reconstructed fluxes are very similar in the different cases.

\subsection{Argon Charged Current Reaction}

 Concerning the reconstruction of the $F_{\rm ArCC}$ 
spectrum of Eq.~(\ref{eq:farcc}) 
 we use the following equation
\begin{equation}
\left.\frac{d{\tilde F}_{\rm ArCC}}{dE_{\nu}}\right\vert_{\bar{E}_i}=\frac{1}{N_{\rm Ar}\sigma^{\rm tot}_{\rm ArCC}(\bar{E}_i)}\frac{N^i_{\rm ArCC}}{E_{\rm vis}^{i}}\,,
\label{inversion_IBD}
\end{equation}
where $\sigma^{\rm tot}_{\rm ArCC}$ is the sum over all excited states of $^{40}$Ar. Here we are neglecting the energy resolution of the detector. Note that we bin the events in order to obtain the neutrino flux for the same $E^i_\nu$ adopted in the case of pES.
The reconstructed fluxes ${\tilde F}_{\rm ArCC}$ and the original ones ${F}_{\rm ArCC} $   are shown in Fig.~\ref{fig:ArCC_inverted}
 for different oscillation scenarios
for W (left panel) and G (right panel) models. While matter effects with IO and flavor equalization are practically degenerate, the matter effects in NO scenario is distinguishable from the other two, especially for the W model.

\section{Identifying the flavor conversion scenario}  
\label{sec:identflav}
The main goal of our work is to assess if one can empirically exclude the possibility that either only matter effects or complete flavor equalization occurs during the accretion phase, without assuming a form for the neutrino energy spectra. We define the ratio of un-oscillated electron and non-electron fluences as 
\begin{equation}
x \equiv \frac{F^0_{\nu_e}}{F^0_{\nu_x}}\lesssim 1\, \qquad \text{and} \qquad \bar{x}\equiv\frac{F^0_{\bar\nu_e}}{F^0_{\nu_x}}\lesssim 1\,.
\end{equation}
These ratios should typically exhibit the hierarchy
\begin{equation}
x\lesssim{\bar x}\lesssim 1
\end{equation}
in the high-energy tails of the (anti)neutrino fluence spectra, because for
$(E>E_c, {\bar E}_c)$ one has the hierarchy $F^0_{\nu_e} < F^0_{\bar\nu_e} < F^0_{\nu_x}$. In the extreme tails, one expects $x,\,\bar{x}\ll1$.
This focus on the high-energy tail is also motivated by the fact that the pES channel has a threshold of approximately 0.2 MeV in visible energy, so that it is only sensitive to (anti)neutrinos in the high-energy tail with larger energies.

The ratios of the neutrino fluences probed by the interaction channels considered here, can be seen to be
\begin{eqnarray}
R= \frac{F_{\rm pES}}{F_{\rm ArCC}}  = \frac{4+x+\bar{x}}{P_{ee}x  + (1-P_{ee})}&=&
 \left\{
\begin{array}{lll}
 \frac{4}{1-P_{ee}}&\quad\quad\, x,\bar{x} \ll 1\\
\frac{4+\bar{x}}{1-P_{ee}}&\quad\quad\, x \ll 1,\text{ and } \bar{x}\lesssim 1\\
6&\quad\quad\, x \lesssim \bar{x} \lesssim 1\\
\end{array}\right.
 \,,
\label{eq:rationu}\\
{\bar R}= \frac{F_{\rm pES}}{F_{\rm IBD{\phantom{x}}}}= \frac{4+x+\bar{x}}{{\bar P}_{ee}\bar{x}  + (1-{\bar P}_{ee})}&=&
 \left\{
\begin{array}{lll}
 \frac{4}{1-{\bar P}_{ee}}&x,\bar{x} \ll 1\\
\frac{4+\bar{x}}{{\bar P}_{ee}\bar{x}+1-{\bar P}_{ee}}&x \ll 1,\text{ and } \bar{x}\lesssim 1\\
6&x \lesssim \bar{x} \lesssim 1\\
\end{array}\right.
 \,.
\label{eq:rationubar}
\end{eqnarray}

\begin{figure}[!t]
\includegraphics[width=0.85\textwidth]{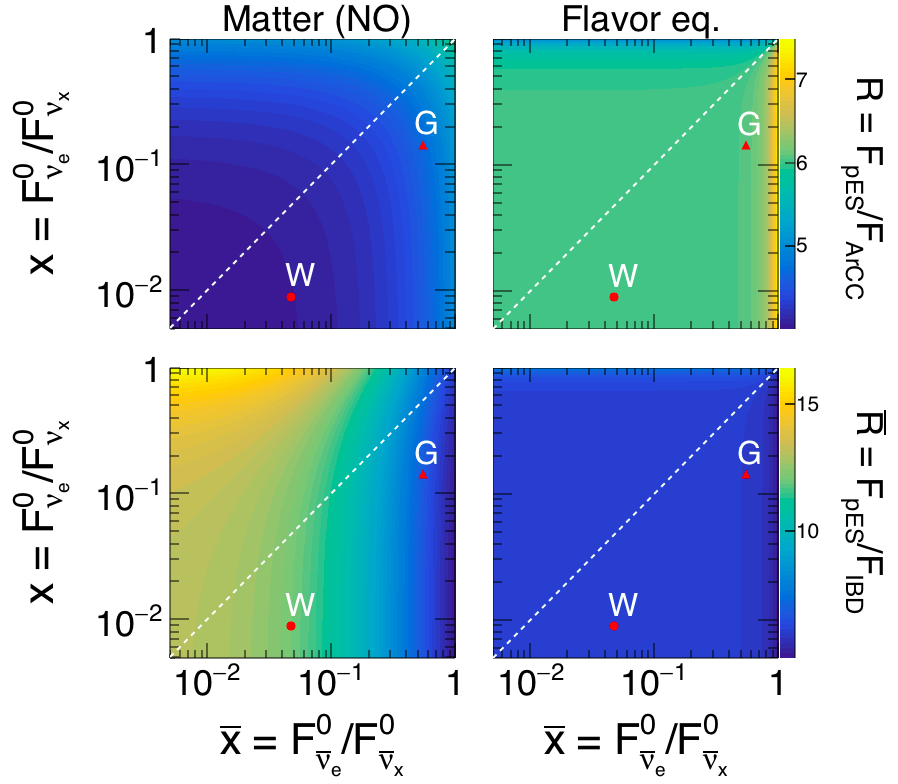}\\
\caption{Ratio $R$ (upper panels) and ${\bar R}$  (lower panels) in NO. Left panels refer to the matter effects only scenario, while
right panels are for complete flavor equalization. The dot refers to Wroclaw model, while the triangle is for  the Garching one. 
The physical region in the high-energy spectral tails is below the dashed line corresponding 
to $x={\bar x}$.
(see text for details)} 
\label{fig:R_NH}
\end{figure}

On the right-hand-sides of Eqs.~(\ref{eq:rationu}) and ({\ref{eq:rationubar}}) we have reported the limiting values of the ratios in the high-energy regime. The ability to distinguish the flavor equalization scenario from pure matter effects depends on the both the differences in the survival probability $P_{ee}$ (see Table \ref{tab:survival}) as well as between the energy spectra of electron and non-electron (anti)neutrino fluxes parametrized through $x$ and $\bar{x}$. In the following, we will develop an oscillation scenario identification scheme by noticing that these ratios cannot take certain values within a specific oscillation scenario (which determines $P_{ee}$ and $\bar{P}_{ee}$) no matter what the energy range (which determines  if $x$, $\bar{x}$ $\lesssim 1$ or $\ll1$).

\subsection{Normal Ordering}

Let us first consider the scenario that neutrinos have a normal mass ordering. Analyzing the ratio $R$, one finds the limiting values,
\begin{equation}
R_{\text{ME}}=
 \left\{
\begin{array}{lll}
4&x,\bar{x} \ll 1\\
5&x \ll 1,\text{ and } \bar{x}\lesssim 1\\
6&x \lesssim \bar{x} \lesssim 1\\
\end{array}\right.
 \,,\qquad
 R_{\text{FE}}=
 \left\{
\begin{array}{lll}
6&x,\bar{x} \ll 1\\
7.5&x \ll 1,\text{ and } \bar{x}\lesssim 1\\
6&x \lesssim \bar{x} \lesssim 1\\
\end{array}\right. \, .
\end{equation}

Since a pure matter effects scenario cannot give $R >6$, we take this value as discrimination
threshold. 
\begin{figure}[!t]
\includegraphics[width=\textwidth]{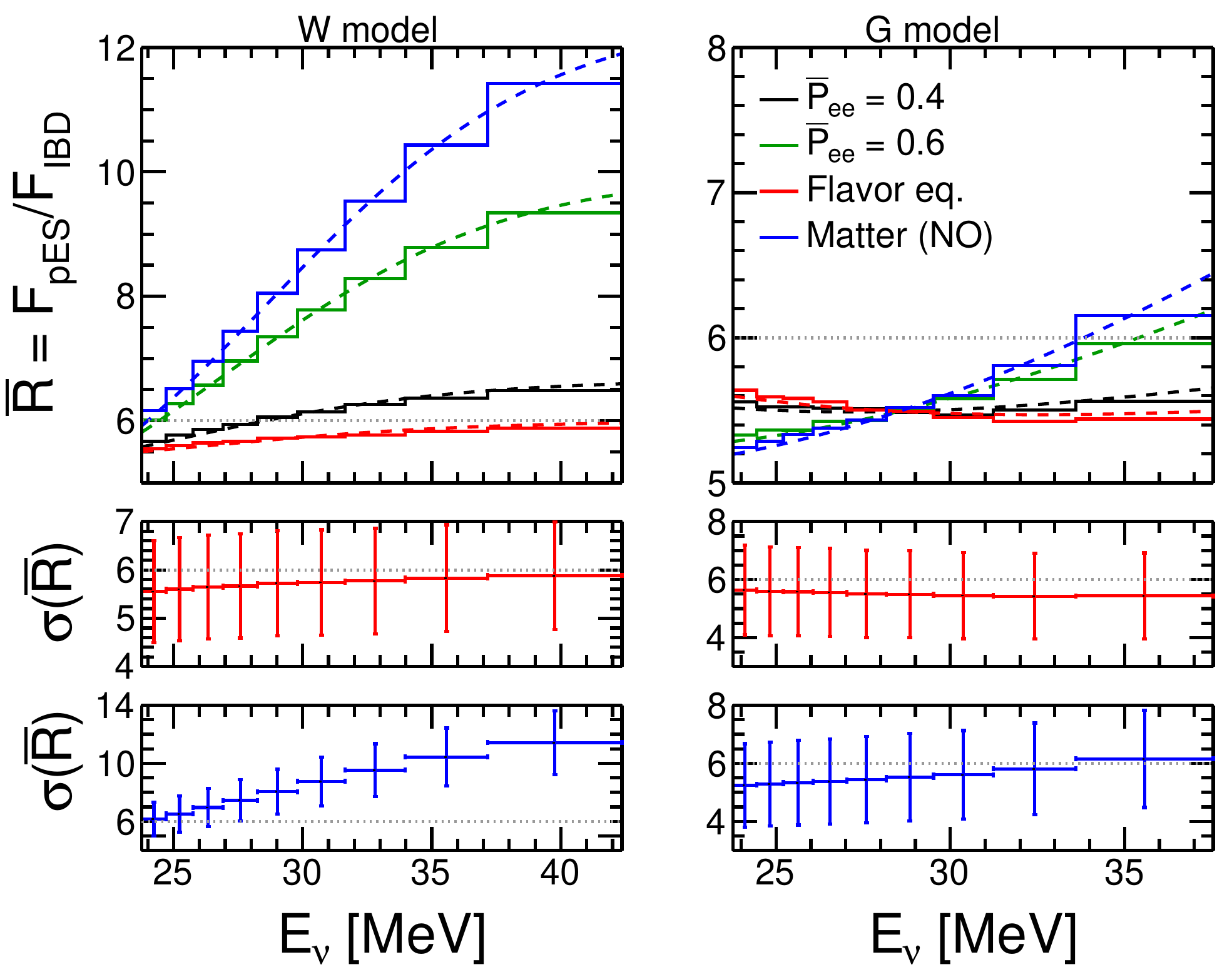}
\caption{Upper panels: Ratio ${\bar R}$ in NO for different values of ${\bar P}_{ee}$. Histograms represent
the results  obtained from the inversion procedure, while dashed curves are obtained calculating ${\bar R}$
directly from the oscillated $\nu$ fluxes. The horizontal dashed line at ${\bar R}=6$ represents our threshold value:
${\bar R}> 6$ excludes complete flavor equalization.  
Middle panels: statistical errors 
for the case of  complete flavor equalization.  Lower panels: statistical errors 
for the case with only matter effects.  Left panels refer to W model, while right panels are for G model.} 
\label{fig:RbarNH}
\vspace{0.5cm}
\end{figure}
\begin{figure}[!t]
\includegraphics[width=\textwidth]{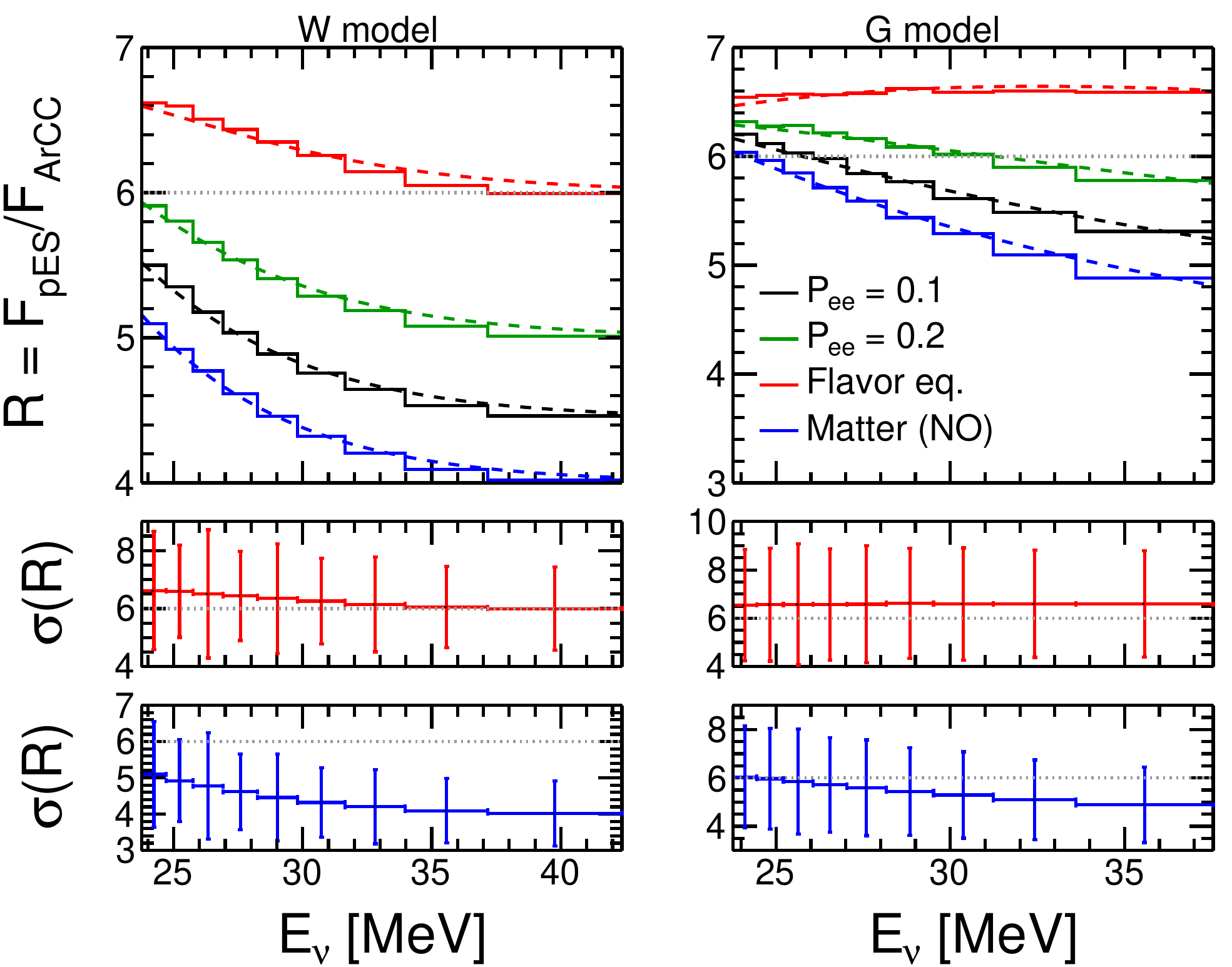}
\caption{Ratio $R$ in NO in the same format as in Fig. \ref{fig:RbarNH}. Values of $R>6$ exclude pure matter effects, while
$R<6$ excludes complete flavor equalization.} 
\label{fig:RNH}
\end{figure}    
From Table~\ref{tab:survival} it is clear that a large difference between the survival probability in the pure matter effects versus the flavor equalization scenario is found also for ${\bar\nu}$, giving
\begin{equation}
\bar{R}_{\text{ME}}=
 \left\{
\begin{array}{lll}
13.3&x,\bar{x} \ll 1\\
5&x \ll 1,\text{ and } \bar{x}\lesssim 1\\
6&x \lesssim\bar{x} \lesssim 1\\
\end{array}\right.
 \,,\qquad
 {\bar R}_{\text{FE}}=
 \left\{
\begin{array}{lll}
6&x,\bar{x} \ll 1\\
5&x \ll 1 ,\text{ and } \bar{x}\lesssim 1\\
6&x \lesssim\bar{x} \lesssim 1\\
\end{array}\right.\, .
\end{equation}
Note that complete flavor equalization cannot give ${\bar R}> 6$, so we take  ${\bar R}=6$ as a conservative discrimination threshold between the matter effects and flavor equalization scenarios.

All possible values of $R$ and $\bar{R}$, as a function of $x$ and $\bar{x}$, are shown in the upper and lower panels of Fig.~\ref{fig:R_NH}, respectively, for the oscillation scenarios with only matter effects (left) and complete flavor equalization (right). The region corresponding to the high-energy tails is below the dashed line $x={\bar x}$. The value of ${\bar R}$ for the W and G models at $E_{\nu}=40$ MeV are also shown using a dot and triangle, respectively. One finds that $R \in [4,13.3]$, while $\bar{R} \in [4,6]$. Values outside this range indicate either a nonstandard scenario (e.g., oscillations into sterile neutrinos, or neutrino decay) or large statistical fluctuations. Thus, one has the following options for NO:
\begin{itemize}
\item $R>6$:  excludes pure matter effects,
\item $R\lesssim 6$:  excludes complete flavor equalization,
\item ${\bar R} > 6$: excludes complete flavor equalization,
\item ${\bar R} \sim 5-6$:  degeneracy between pure matter effects and complete or partial flavor equalization, which might be removed in combination with $R$.
\end{itemize}

\begin{figure}[!t]
\includegraphics[width=\textwidth]{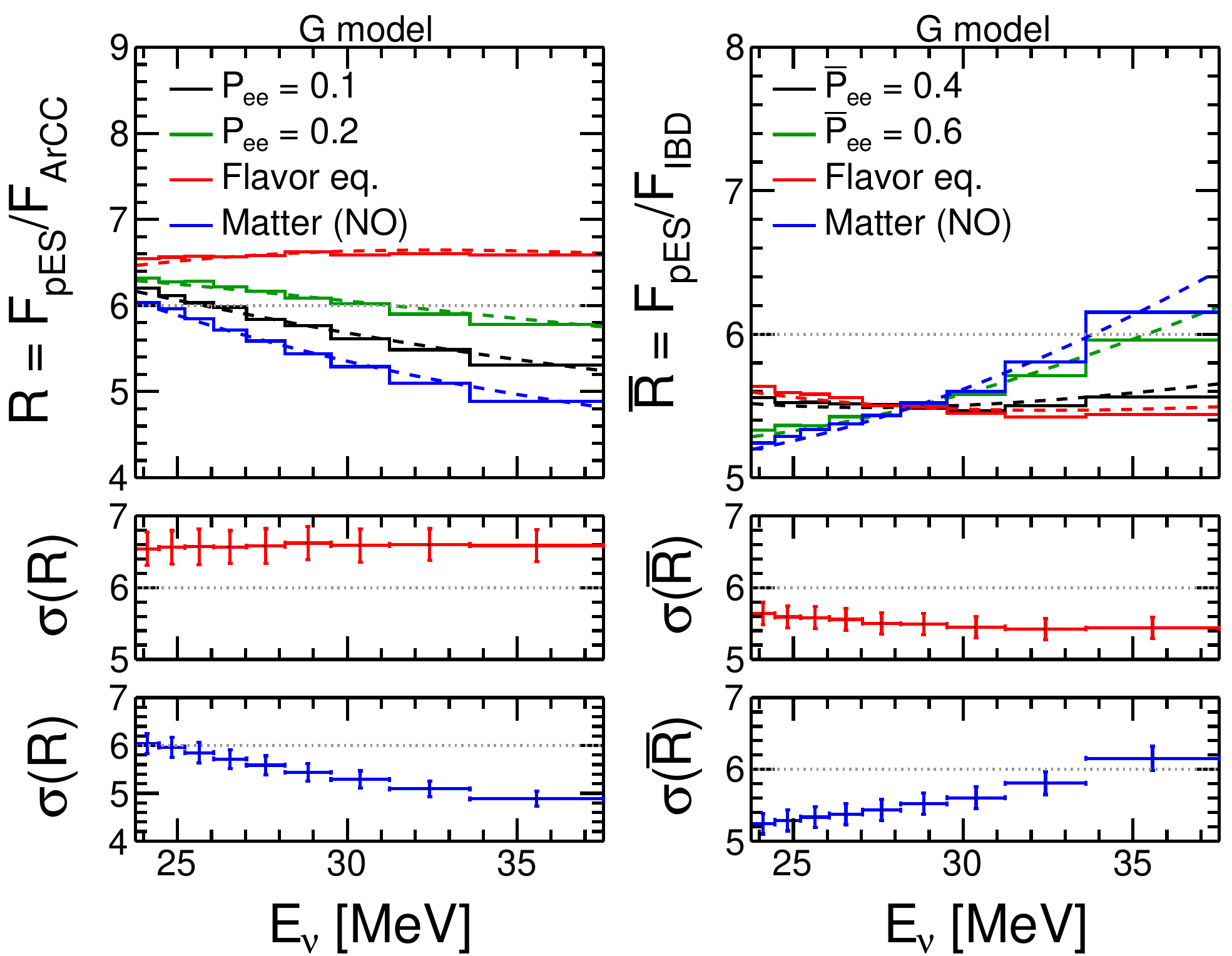}
\caption{Ratio of events in NO for Garching model for a SN at $d=1$~kpc.  Upper panels: Ratio ${R}$ (left) and  ${\bar R}$ (right) for different values of ${\bar P}_{ee}$. 
Middle panels: statistical errors 
for the case of  complete flavor equalization.  Lower panels: statistical errors 
for the case of pure matter effects.} 
\label{fig:Rgar1}
\end{figure}  

From Fig.~\ref{fig:R_NH} it becomes apparent that  the discrimination between the two oscillation scenarios would ideally be possible for both 
W and G models; more easily for the W model (${\bar x} \simeq 5 \times 10^{-2}$) than for the G model (${\bar x}\simeq 0.5 $). However, in order to perform a more realistic investigation, we now evaluate the ratio ${\bar R}$ using the fluxes ${\tilde F}$ obtained from the inversion process described in Sec.~\ref{sec:recon}.
 In Fig.~\ref{fig:RbarNH} we plot ${\bar R}$  for different values of ${\bar P}_{ee}$ (upper panels).
Left panels refer to W model, while right panels are for G model.
Continuous lines represent
${\bar R}$ obtained through the inversion procedure, while dashed curves are for the true oscillated $\nu$ fluxes. The horizontal dashed line 
${\bar R}= 6$
represents our threshold value.
One sigma statistical errors are shown in the  middle panels 
for the case of  complete flavor equalization, and in the lower panels
for matter effects only. A conservative estimate of the statistical significance of one of the options described before is given by simply evaluating the distance of $R$ or $\bar{R}$ in the last energy bin from the correspondent threshold value in terms of  number of sigmas. Considering more energy bins will lead, in some cases, to higher statistical significance. However, since a priori we do not know the critical energy $E_{c}$, focusing only on the highest energy bin is the safest option to avoid a wrong evaluation of the significance.
For the W model, in the case of only matter effects we obtain ${\bar R} \simeq 11.5$, which is enough to disfavour complete flavor equalization at $\gtrsim2\sigma$.  
Note that cases of partial flavor equalization would lead to $0.33 <{\bar P}_{ee} < 0.7$. These would be distinguishable from complete flavor equalization (${\bar P}_{ee}=0.33$)  if ${\bar R} \gg 6$  (e.g., ${\bar P}_{ee}=0.6$ in the figure). 
On the other hand, for the G model, we find ${\bar R} \sim 5.5
$ for flavor equalization, and ${\bar R} \sim 6.5$ for only matter effects. These do not show large deviations with respect to ${\bar R} =6$, and thus
no discrimination  between them is possible within the statistical error.

In Fig.~\ref{fig:RNH} we show the reconstructed $R$
for the two benchmark models, in the same format of 
Fig.~\ref{fig:RbarNH}. Considering the W model, we obtain $R\simeq 4$ when only matter effects occurs,
which is enough to disfavour complete flavor equalization at $\sim2\sigma$. Conversely, complete flavor equalization would not show  sizeable deviations with respect 
to ${ R} \simeq 6$, precluding the exclusion of a pure matter effects scenario.
Assuming the G model we obtain $R\simeq 6$ within the error bars in all oscillation scenarios, preventing the exclusion of any of them.

Altogether, one can exploit the two ratios to acquire some information on the oscillation scenario if the spectra are not too similar. In particular, if ${\bar R}>6$ and ${ R} < 6$ one would exclude a complete flavor equalization scenario. Whereas if $R>6$, with significant confidence, it excludes pure matter effects. Obviously the power of our method increases when considering a SN distance smaller than $d=10$ kpc, as a result of reduced error bars. In this case, also the G model might be enough to discriminate  among different oscillation scenarios. This is evident from Fig.~\ref{fig:Rgar1}, which shows $R$ and $\bar{R}$ for the G model and $d=1$ kpc. For instance the values of $R$ are always above (below) the threshold for flavor equalization (matter effects) even considering the statistical errors. For instance, if FE (matter effects in NO) occurred in the supernova we could exclude the other flavor scenario with a significance of $\sim3\sigma$ ($>5\sigma$).

\begin{figure}[!t]
\includegraphics[width=0.85\textwidth]{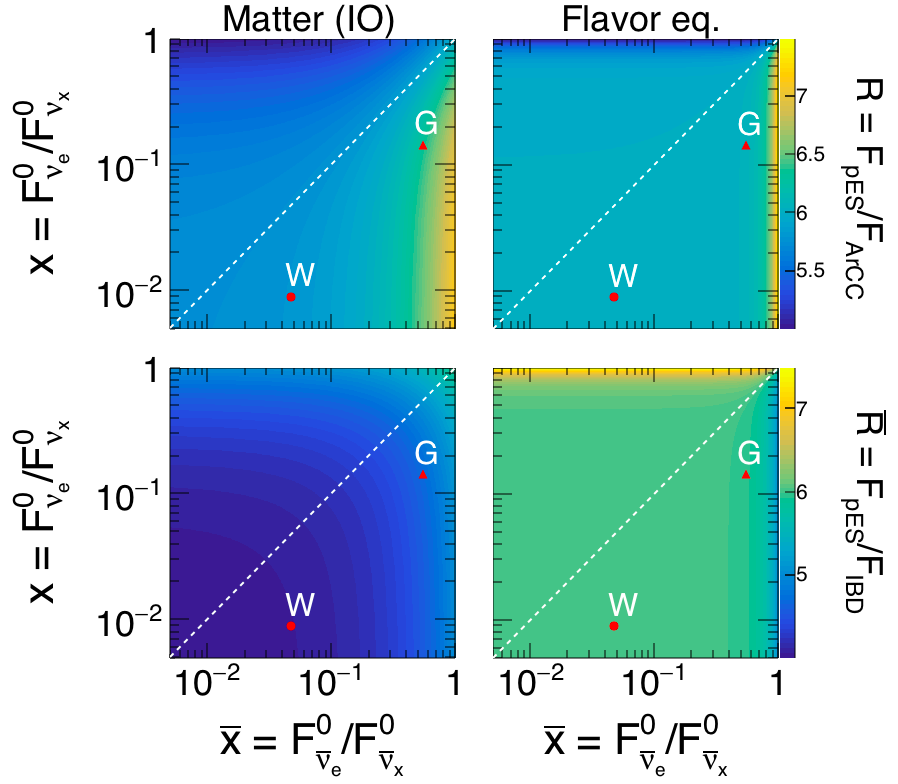}
\caption{Ratio $R$ (upper panels) and ${\bar R}$  (lower panels) in IO. Left panels refer to the pure matter effects scenario, while
right panels are for complete flavor equalization. The dot refer to Wroclaw model, while the triangle is for  the Garching one. 
The physical region in the high-energy spectral tails is below the dashed line corresponding 
to $x={\bar x}$.
(see text for details)} 
\label{fig:R_IH}
\end{figure}

\subsection{Inverted Ordering}

We consider now the case that neutrinos have an inverted mass ordering. As discussed in Sec.\,\ref{sec:osciscen}, in this case the neutrino channel is not useful 
to disentangle the oscillation scenarios, since the survival probabilities are very similar for flavor equalization and matter effects
(see Table \ref{tab:survival}). This is evident from the ratio $R$ in the upper panels of Fig.~\ref{fig:R_IH}, which look essentially identical. Therefore, we focus on the ratio ${\bar R}$. In this case,
\begin{equation}
\bar{R}_{\text{ME}}=
 \left\{
\begin{array}{lll}
4&x,\bar{x} \ll 1\\
5&x \ll 1,\text{ and } \bar{x}\lesssim 1\\
6&x \lesssim\bar{x} \lesssim 1\\
\end{array}\right.
 \,,\qquad
 {\bar R}_{\text{FE}}=
 \left\{
\begin{array}{lll}
6&x,\bar{x} \ll 1\\
5&x \ll 1 ,\text{ and } \bar{x}\lesssim 1\\
6&x \lesssim\bar{x} \lesssim 1\\
\end{array}\right.\, .
\end{equation}
The ratio ${\bar R}$ for these two oscillation scenarios is shown in the lower panels of Fig.~\ref{fig:R_IH}.
As both the scenarios give ${\bar R} \leq 6$, this time 
we take as discrimination threshold value ${\bar R}=5$,
finding the following options
\begin{itemize}
\item ${\bar R} <5$: excludes complete flavor equalization,
\item ${\bar R}>5$: degeneracy of scenarios.
\end{itemize}

\begin{figure}[!t]
\includegraphics[width=\textwidth]{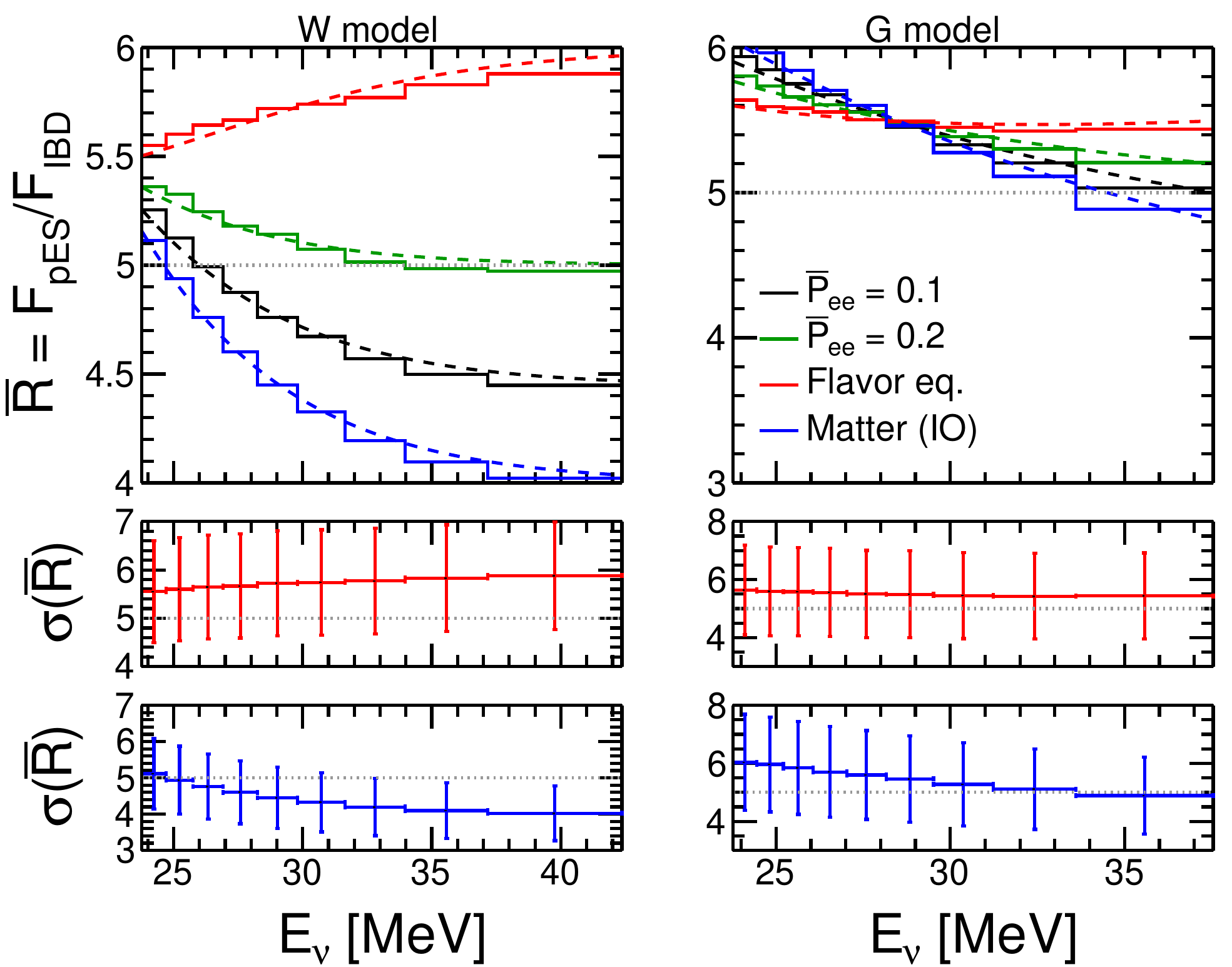}
\caption{Ratio ${\bar R}$ in IO in the same format as in Fig.~\ref{fig:RbarNH}. In this case our threshold value is 
${\bar R}=5$.  ${\bar R}<5$ excludes complete flavor equalization.} 
\label{fig:RbarIH}
\end{figure}

From Fig.~\ref{fig:R_IH}  it seems that in principle the two oscillation scenarios could be discriminated for  
W model, if neutrino fluxes were perfectly known, whereas the differences between the two scenarios are modest for the G model. 
We show the reconstructed ${\bar R}$ for the two benchmark models in Fig.~\ref{fig:RbarIH}, in the same format of 
Fig.~\ref{fig:RbarNH}. 
We see that for the W model, pure matter effects leads to  ${\bar R} \simeq 4$ allowing one to have an hint for  complete flavor equalization at $\sim1\sigma$.
Conversely, for the G model both matter effects and flavor equalization give ${\bar R}>5$ in the highest energy, precluding any discrimination. 
 
\vspace{0.5cm}
\section{Summary and conclusions}
\label{sec:summary}

The characterization of flavor conversions in SN neutrinos is still an unsettled problem and perhaps a definite
answer will not be available in the immediate future. Therefore, it is important to investigate the possibility  to exploit a  future observation of a Galactic SN neutrino burst 
 to discriminate among possible oscillation scenarios. 
In this context, we focussed on the SN accretion phase where two alternative oscillation scenarios might be realized, namely
a pure matter effects or flavor equalization among different neutrino spectra. 
In order to probe these two scenarios we proposed  a fit-free analysis at large next-generation underground detectors
(i.e., 0.4 Mton water Cherenkov, 20 kton scintillation, and 40 kton liquid Argon detectors).
Our method uses the pES in a scintillator detector, which is insensitive to oscillation effects, as a benchmark channel.
We compared events in this channel with events due to IBD of $\bar\nu_e$ in a water Cherenkov detector and with events from the charged-current interactions of $\nu_e$ in Argon detector. We have shown that the ratio $R=F_{\rm pES}/F_{\rm ArCC}$ as well as $\bar{R}=F_{\rm pES}/F_{\rm IBD}$ might allow one to exclude the possibility of either complete flavor equalization or of pure matter effects in some cases. To summarize our scheme, for normal neutrino mass ordering,
\begin{itemize}
\item $R>6$:  excludes pure matter effects,
\item $R\lesssim 6$:  excludes complete flavor equalization,
\item ${\bar R} > 6$: excludes complete flavor equalization,
\item ${\bar R} \sim 5-6$:  degeneracy between matter effects and complete or partial flavor equalization, which might be removed in combination with $R$.
\end{itemize} 
Whereas, for inverted neutrino mass ordering,
\begin{itemize}
\item ${\bar R} <5$: excludes complete flavor equalization,
\item ${\bar R}>5$: degeneracy of scenarios.
\end{itemize}
 
We have considered two benchmark SN neutrino models as specific test-cases. For the W model, which has larger spectral differences among the different flavors, our strategy shows a stronger discriminatory power than with the G model, where $\nu$ fluxes are more similar. Indeed, the efficacy of the proposed  method mostly depends on the hierarchy among the neutrino fluxes. In this context, current SN simulations have improved in many ways relative to what was state-of-the-art one or two decades ago. Nevertheless, their predictions should still be considered as indicative. It is important to perform direct  empirical tests, and if our method gives a positive indication it will also indicate  sizeable differences in the flavor spectra.
 
We stress that our procedure is really model-independent, in the sense that it does not rely fitting or modelling the spectral shape of the neutrino fluences. It is also not affected by the uncertainties in the characterization of the high-energy tails of the SN neutrino spectra, as long as $F^0_{\nu_e,\bar{\nu}_e}\lesssim F^0_x$. It is however based on a direct reconstruction on the neutrino fluxes of the observed events rate, and could be improved by
including the effects of finite energy resolution and limited detector acceptance in the reconstruction procedure. In order to take into account these effects, an unfolding procedure has been recently applied to SN neutrino spectra in~\cite{Li:2017dbg}. Furthermore, our analysis might be improved combining it with the rise-time analysis of the neutrino signal during the accretion phase~\cite{Serpico:2011ir}, and perhaps use of subleading detection channels to gain information on neutrino spectra, especially $\nu_e$~\cite{Laha:2013hva, Laha:2014yua, Nikrant:2017nya,GalloRosso:2017mdz}.
Finally,  it would be interesting to extend our studies to other classes of SNe,
e.g., electron capture  or failed SNe where one would expect a different neutrino flux
hierarchy and different oscillation effects during the accretion phase.
 We leave all these improvements of our method for a future work.
 
The strategy developed in our work confirms the high physics potential of SN neutrino detection in shedding light on flavor conversion effects occurring in the deepest stellar regions. If Nature is kind, such a detection would allow us to get a direct evidence of the unusual behavior of the dense SN neutrino gas. If hints of flavor equalization are found, they will dramatically change the current paradigm of SN explosion and of stellar nucleosynthesis. We hope that this exciting perspective will motivate further studies in this direction.

\section*{Acknowledgments}

We are grateful to John Beacom, Takatomi Yano, Eligio Lisi, Georg Raffelt, and Irene Tamborra for useful comments on the manuscript. A.M. and F.C. acknowledge kind hospitality at CERN (Geneva) and at TIFR (Mumbai) were this work was initiated. The work of B.D. is partially supported by the Dept. of Science and Technology of the Govt.\,of India through a Ramanujan Fellowship and by the Max-Planck-Gesellschaft through a Max-Planck-Partnergroup. The work of A.M. is supported by the Italian Istituto Nazionale di Fisica Nucleare (INFN) through the ``Theoretical Astroparticle Physics'' project and by Ministero dell'Istruzione, Universit\`a e Ricerca (MIUR). The work of F.C. is supported partially by the Deutsche Forschungsgemeinschaft through Grant No. EXC 153 (Excellence Cluster ``Universe") and Grant No. SFB 1258 (Collaborative Research Center ``Neutrinos, Dark Matter, Messengers") as well as by the European Union through Grant No. H2020-MSCA-ITN-2015/674896 (Innovative Training Network ``Elusives").
 
\clearpage 

\bibliography{biblio}
  
\end{document}